# Logical Foundations of RDF(S) with Datatypes

**Jos de Bruijn** BRUIJN@KR.TUWIEN.AC.AT
**Stijn Heymans** HEYMANS@KR.TUWIEN.AC.AT
*Vienna University of Technology*
*Favoritenstraße 9-11, A-1040 Vienna, Austria*

## Abstract

The Resource Description Framework (RDF) is a Semantic Web standard that provides a data language, simply called RDF, as well as a lightweight ontology language, called RDF Schema. We investigate embeddings of RDF in logic and show how standard logic programming and description logic technology can be used for reasoning with RDF. We subsequently consider extensions of RDF with datatype support, considering $D$ entailment, defined in the RDF semantics specification, and $D^*$ entailment, a semantic weakening of $D$ entailment, introduced by ter Horst. We use the embeddings and properties of the logics to establish novel upper bounds for the complexity of deciding entailment. We subsequently establish two novel lower bounds, establishing that RDFS entailment is PTime-complete and that simple-$D$ entailment is coNP-hard, when considering arbitrary datatypes, both in the size of the entailing graph. The results indicate that RDFS may not be as lightweight as one may expect.

## 1. Introduction

The Resource Description Framework (RDF) (Klyne & Carroll, 2004), together with its vocabulary description language RDF Schema (RDFS) (Brickley & Guha, 2004), constitutes the basic language of the Semantic Web. Statements in RDF are triples of the form $\langle s, p, o\rangle$. Sets of triples are called *RDF graphs*: intuitively, each triple can be viewed as an edge from node $s$ to node $o$ with label $p$. Here, $s$, $o$, and $p$ are constant symbols – uniform resource identifiers (URIs) or literals (e.g., strings) – or anonymous identifiers, called *blank nodes*. Consider, for example, the graphs $S = \{\langle o, \texttt{rdf:type}, A\rangle, \langle A, \texttt{rdfs:subClassOf}, B\rangle\}$ and $E = \{\langle o, \texttt{rdf:type}, B\rangle\}$. Hayes (2004) defines notions of RDF and RDFS entailment. We have that, compared with RDF entailment, RDFS entailment gives additional meaning to `rdfs:subClassOf` statements: $S$ RDFS-entails $E$, but $S$ does not RDF-entail $E$.

The RDF semantics specification (Hayes, 2004) defines four increasingly expressive normative entailment relations between RDF graphs, namely *simple*, *RDF*, *RDFS*, and *D entailment*, where the latter extends RDFS entailment with support for datatypes (e.g., strings and integers). Furthermore, it defines *extensional RDFS* (eRDFS) entailment as a possible extension of RDFS entailment that is more in line with description logic-based languages such as OWL DL (Patel-Schneider, Hayes, & Horrocks, 2004) and OWL 2 DL (Motik, Patel-Schneider, & Parsia, 2009b). Intuitively, the difference between the RDFS and eRDFS entailment regimes is that, for the latter, whenever an ontological relation (e.g., subclass or property domain) implicitly holds in an interpretation, the corresponding RDF statement (`rdfs:subClassOf`, `rdfs:domain`, respectively) must be true, whereas this is not always the case with the RDFS entailment regime. The following example illustrates this difference.

**Example 1.** *Let S be the graph*

$$\{\langle mother, \texttt{rdfs:subPropertyOf}, parent\rangle, \langle parent, \texttt{rdfs:domain}, Person\rangle\}$$

*which says that Person is in the domain of parent, and the property mother is a subproperty of parent. Using eRDFS entailment we can conclude from S that Person is in the domain of mother:*

$$S \models_{erdfs} \langle mother, \texttt{rdfs:domain}, Person\rangle$$

*since it must the case that the subject of any mother triple has the type Person; thus, Person is implicitly in the domain of mother. We cannot draw this conclusion when using RDFS entailment; in RDFS, only explicitly asserted domain constraints can be derived.*

We further also consider $D^*$ entailment (ter Horst, 2005), which is a semantic weakening of $D$ entailment for the purpose of more efficient computation of consequences. $D^*$ entailment extends RDFS entailment, but is not more expensive in terms of computational complexity.

There have been several investigations into the formal properties of the RDF semantics (Gutierrez, Hurtado, & Mendelzon, 2004; Gutierrez, Hurtado, Mendelzon, & Pérez, 2010; de Bruijn, Franconi, & Tessaris, 2005; ter Horst, 2005): Gutierrez et al. (2004, 2010) reconstruct the semantics from a graph database perspective, and de Bruijn et al. (2005) reconstruct the semantics from a logical language perspective. The investigation of the RDF semantics by ter Horst (2005) stays very close to the RDF specification. Additionally, ter Horst shows that the entailment rules for computing RDFS entailment presented in the original specification (Hayes, 2004) are not complete with respect to the RDFS semantics. These reconstructions have led to a number of complexity results for RDF entailment. In particular, simple, RDF, and RDFS entailment are NP-complete in the combined size of the graphs. This high complexity is due to the presence of blank nodes (essentially existentially quantified variables): if the entailed graph is known to be ground, the respective problems turn out to be decidable in polynomial time. These bounds have not been shown to be tight. As we will show in Section 5, the bound is tight for RDFS entailment, but not for simple and RDF entailment, which can be decided in logarithmic space.

To investigate the relationship between RDF and logic we embed the various RDF entailment regimes in F-Logic (Kifer, Lausen, & Wu, 1995), which is a syntactic extension of first-order logic (FOL) with object oriented modeling constructs. F-Logic has constructs to explicitly specify attributes, as well as generalization/specialization and instantiation relationships. Like RDFS, the syntax of F-Logic has some seemingly higher-order features, namely, the same identifier can be used for a class, an instance, and an attribute. However, the semantics of F-Logic is strictly first-order (Kifer et al., 1995). It turns out that the attribute value construct in F-Logic is exactly equivalent to the triple construct in RDF, and the typing (class membership) construct in F-Logic is very close in spirit to the one in RDF.

In addition, we consider the embedding of a large subset of extensional RDFS in FOL and the tractable description logic language $DL\text{-}Lite_{\mathcal{R}}$ (Calvanese, Giacomo, Lembo, Lenzerini,

& Rosati, 2007), thereby showing that, under certain restrictions, extensional RDFS can be seen as a standard first-order knowledge representation language.

Our contributions with this paper can be summarized as follows.

1. We define embeddings of simple, RDF, RDFS, and extensional RDFS into F-Logic, and show that simple, RDF, and RDFS entailment can be decided using standard logic programming techniques, as their embeddings are in the Horn fragment of F-Logic.

2. We define an alternative, direct embedding of extensional RDFS into the Horn fragment of F-Logic for a fragment of RDF graphs, namely those in which the RDFS vocabulary is only used in a *standard* way. We subsequently exploit earlier results about the relationship between F-Logic statements and description logic statements (de Bruijn & Heymans, 2008) to show that extensional RDFS reasoning with ground RDF graphs can be reduced to reasoning in the tractable description logic $DL\text{-}Lite_{\mathcal{R}}$ (Calvanese et al., 2007).

3. We extend the embeddings mentioned under 1. with support for datatypes, considering both $D^*$ and $D$ entailment. The embeddings of the extensions of simple, RDF, and RDFS entailment with $D^*$ datatype support are all essentially in the Horn fragment of F-Logic. The extensions of simple, RDF, and RDFS with $D$ datatype support can be embedded in the Horn fragment of F-Logic when suitably restricting the datatypes that may be considered.

4. We analyze the complexity of deciding the mentioned entailment relations. From the mentioned embeddings we obtain a number of novel complexity upper bounds, namely, simple and RDF entailment, as well as their extensions with datatypes (under suitable restrictions), are in LogSpace in the size of the entailing graph and a large fragment of extensional RDFS entailment is in NP in the combined size of the graphs and in PTime in the size of the entailing graph. We also establish a novel PTime lower bound for RDFS entailment and a novel coNP lower bound for simple entailment extended with $D$ datatype support, when considering arbitrary datatypes, both in the size of the entailing graph. See Table 2 on page 553 for an overview of the complexity results for RDF.

The structure of the remainder of the paper is as follows. In Section 2 we review the logics under consideration, namely F-Logic and $DL\text{-}Lite_{\mathcal{R}}$. In Section 3 we review the RDF(S) semantics, define embeddings into F-Logic and FOL, show faithfulness of these embeddings, and demonstrate the relationship with $DL\text{-}Lite_{\mathcal{R}}$. In Section 4 we consider extensions of the RDF entailment regimes with datatype support based on both $D^*$ and $D$ entailment and embeddings of these extensions into logic. In Section 5 we extensively investigate the complexity of the various RDF entailment regimes. We conclude the paper and outline future work in Section 6.

This paper extends a paper we published at the International Semantic Web Conference (de Bruijn & Heymans, 2007) with embeddings of the $D^*$ and $D$ entailment regimes and novel lower bounds for the complexity of deciding RDFS, $D^*$, and $D$ entailment.

For reasons of legibility, the definitions of the various RDF-related notions of interpretation may be found in Appendix A, the embeddings of the RDF entailment regimes may be found in Appendix B, and the proofs of Sections 3 and 4 may be found in Appendix C.

## 2. Preliminaries

In this section we review F-Logic and $DL\text{-}Lite_{\mathcal{R}}$.

### 2.1 Frame Logic

We consider Frame Logic (F-Logic) as defined by Kifer, Lausen, and Wu (1995). To simplify matters, and because these constructs are not necessary for the embedding of RDF, we do not consider function symbols, parameterized methods, functional (single-valued) methods, inheritable methods, and compound molecules, following de Bruijn and Heymans (2008).

The signature of an F-language $\mathcal{L}$ is of the form $\Sigma = \langle \mathcal{C}, \mathcal{P} \rangle$ with $\mathcal{C}$ and $\mathcal{P}$ disjoint sets of constant and predicate symbols; each predicate symbol has an associated arity $n \geq 0$. Let $\mathcal{V}$ be a set of variable symbols. Terms and atomic formulas are constructed as usual: $x \in \mathcal{V}$ and $c \in \mathcal{C}$ are terms and $\top$, $\bot$, $p(t_1, \ldots, t_n)$, and $t_1 = t_2$ are atomic formulas, with $p \in \mathcal{P}$ an $n$-ary predicate symbol, and $t_1, \ldots, t_n$ terms.

A molecule in F-Logic is one of the following statements: (i) an *is-a* assertion of the form $t_1 : t_2$, which states that an individual $t_1$ is of type $t_2$, or (ii) a *data molecule* (called "method" by Kifer et al., 1995) of the form $t_1[t_2 \twoheadrightarrow t_3]$, with $t_1$, $t_2$, and $t_3$ terms, which states that an individual $t_1$ has an attribute $t_2$ with value $t_3$. An F-Logic molecule is *ground* if it does not contain variables.

Formulas of an F-language $\mathcal{L}$ are either atomic formulas, molecules, or compound formulas which are constructed in the usual way from atomic formulas, molecules, and the logical connectives $\neg, \wedge, \vee, \supset$, the quantifiers $\exists, \forall$ and the auxiliary symbols '(' and ')'. We denote universal closure, i.e., the universal quantification of every variable that has a free occurrence in the formula, with $(\forall)$.

A theory is a set of formulas. A theory or formula is called *equality-free* if the equality symbol '=' does not appear in it.

F-Logic Horn formulas are of the form $(\forall) B_1 \wedge \ldots \wedge B_n \supset H$, with $B_1, \ldots, B_n$, and $H$ atomic formulas or molecules. F-Logic Datalog formulas are F-Logic Horn formulas such that every variable in $H$ occurs in some equality-free $B_i$. The latter condition is called *safeness.*

An *F-structure* is a tuple $\mathbf{I} = \langle U, \in_U, \mathbf{I}_C, \mathbf{I}_{\twoheadrightarrow}, \mathbf{I}_P \rangle$, where $U$ is a non-empty set and $\in_U$ is a binary relation over $U$. A constant symbol $c \in \mathcal{C}$ is interpreted as an element of the domain: $\mathbf{I}_C(c) \in U$. An $n$-ary predicate symbol $p \in \mathcal{P}$ is interpreted as a relation over the domain $U$: $\mathbf{I}_P(p) \subseteq U^n$. $\mathbf{I}_{\twoheadrightarrow}$ associates a binary relation over $U$ with each $k \in U$: $\mathbf{I}_{\twoheadrightarrow}(k) \subseteq U \times U$. Variable assignments $B$ are defined in the usual way.

Given an F-structure $\mathbf{I}$, a variable assignment $B$, and a term $t$ of $\mathcal{L}$, $t^{\mathbf{I},B}$ is defined as: $x^{\mathbf{I},B} = x^B$ for variable symbol $x$ and $t^{\mathbf{I},B} = \mathbf{I}_C(t)$ for $t \in \mathcal{C}$.

Satisfaction of atomic formulas and molecules $\phi$ in $\mathbf{I}$, given a variable assignment $B$, denoted $(\mathbf{I}, B) \models_{\mathsf{f}} \phi$, is defined as
– $(\mathbf{I}, B) \models_{\mathsf{f}} \top$, $\quad (\mathbf{I}, B) \not\models_{\mathsf{f}} \bot$,

– $(\mathbf{I}, B) \models_\mathsf{f} p(t_1, \ldots, t_n)$ iff $(t_1^{\mathbf{I},B}, \ldots, t_n^{\mathbf{I},B}) \in \mathbf{I}_P(p)$,
– $(\mathbf{I}, B) \models_\mathsf{f} t_1 = t_2$ iff $t_1^{\mathbf{I},B} = t_2^{\mathbf{I},B}$,
– $(\mathbf{I}, B) \models_\mathsf{f} t_1 : t_2$ iff $t_1^{\mathbf{I},B} \in_U t_2^{\mathbf{I},B}$, and
– $(\mathbf{I}, B) \models_\mathsf{f} t_1[t_2 \twoheadrightarrow t_3]$ iff $\langle t_1^{\mathbf{I},B}, t_3^{\mathbf{I},B} \rangle \in \mathbf{I}_\twoheadrightarrow(t_2^{\mathbf{I},B})$.

This extends to arbitrary formulas in the usual way. An F-structure $\mathbf{I}$ *satisfies* a formula $\phi$, denoted $\mathbf{I} \models_\mathsf{f} \phi$, if $(\mathbf{I}, B) \models_\mathsf{f} \phi$ for every variable assignment $B$. $\mathbf{I}$ satisfies a theory $\Phi \subseteq \mathcal{L}$ if it satisfies all formulas in $\Phi$; in this case, $\mathbf{I}$ is called a *model* of $\Phi$. A theory $\Phi$ *F-entails* a formula $\phi \in \mathcal{L}$, denoted $\Phi \models_\mathsf{f} \phi$, iff for every model $\mathbf{I}$ of $\Phi$, $\mathbf{I} \models_\mathsf{f} \phi$.

A Herbrand F-structure is an F-structure $\mathbf{I} = \langle U, \in_U, \mathbf{I}_C, \mathbf{I}_\twoheadrightarrow, \mathbf{I}_P \rangle$ such that $U$ is the set of constants and for every constant symbol $c \in \mathcal{C}$, $\mathbf{I}_C(c) = c$. As an abuse of notation, for Herbrand structures we use $\mathbf{I}$ to denote both the structure and the set of ground atomic formulas satisfied by the structure. Finally, a Herbrand F-structure $\mathbf{I}$ is a *minimal* Herbrand model of a theory $\Phi$ if it is a model and there is no Herbrand F-structure $\mathbf{I}'$ that is a model of $\Phi$ such that $\mathbf{I}' \subsetneq \mathbf{I}$.

*Classical first-order logic* (FOL) is F-Logic without molecules. *Contextual first-order logic* is classical FOL where $\mathcal{C}$ and $\mathcal{P}$ are not required to be disjoint, predicate symbols do not have an associated arity, and for every structure $\mathbf{I} = \langle U, \in_U, \mathbf{I}_C, \mathbf{I}_\twoheadrightarrow, \mathbf{I}_P \rangle$, $\mathbf{I}_P$ assigns a relation $\mathbf{I}_P^i(p) \subseteq U^n$ to every $p \in \mathcal{P}$, for every nonnegative integer $i$. We denote satisfaction and entailment in classical and contextual first-order logic using the symbols $\models$ and $\models_c$, respectively. Contextual FOL is sometimes also referred to as FOL with "punning".

F-Logic can be straightforwardly embedded into FOL, as shown in (Kifer et al., 1995, Theorem 18.1).

**Proposition 1.** *Let $\Phi$ and $\phi$ be an F-Logic theory and formula that do not contain the binary and ternary predicate symbols isa and data, respectively, and let $\Phi'$ and $\phi'$ be the FOL theory and formula obtained from $\Phi$ and $\phi$ by replacing every is-a molecule $a : b$ with $isa(a, b)$ and every data molecule $a[b \twoheadrightarrow c]$ with $data(a, b, c)$. Then,*

$$\Phi \models_\mathsf{f} \phi \textit{ iff } \Phi' \models \phi'$$

### 2.2 *DL-Lite*$_\mathcal{R}$

A *DL-Lite*$_\mathcal{R}$ (Calvanese et al., 2007) language consists of pairwise disjoint sets of concept ($N_\mathcal{C}$), role ($N_\mathcal{R}$), and individual ($N_\mathcal{I}$) identifiers. Concepts and roles in *DL-Lite*$_\mathcal{R}$ are defined as follows:

$$\begin{aligned} C_l &\longrightarrow A \mid \exists R \\ C_r &\longrightarrow A \mid \exists R \mid \neg A \mid \neg \exists R \\ R, R' &\longrightarrow P \mid P^- \end{aligned}$$

with $A \in N_\mathcal{C}$ and $P \in N_\mathcal{R}$, $C_l$ and $C_r$ left- (resp., right-)hand side concepts, and $R$ and $R'$ roles.

A *DL-Lite*$_\mathcal{R}$ knowledge base $\mathcal{K} = (\mathcal{T}, \mathcal{A})$ consists of a TBox $\mathcal{T}$, which is a set of inclusion axioms of the forms

$$C_l \sqsubseteq C_r \qquad R \sqsubseteq R'$$

| DL syntax | FOL syntax |
|---|---|
| $\pi(A, X)$ | $A(X)$ |
| $\pi(P, X, Y)$ | $P(X, Y)$ |
| $\pi(P^-, X, Y)$ | $P(Y, X)$ |
| $\pi(\exists R, X)$ | $\exists y(\pi(R, X, y))$ |
| $\pi(\neg A, X)$ | $\neg\pi(A, X)$ |
| $\pi(\neg\exists R, X)$ | $\neg\exists y(\pi(R, X, y))$ |
| $y$ is a new variable | |

| DL syntax | FOL syntax |
|---|---|
| $\pi(C_l \sqsubseteq C_r)$ | $\forall x(\pi(C_l, x) \supset \pi(C_r, x))$ |
| $\pi(R_1 \sqsubseteq R_2)$ | $\forall x, y(\pi(R_1, x, y) \supset \pi(R_2, x, y))$ |
| $\pi(A(a))$ | $A(a)$ |
| $\pi(P(a_1, a_2))$ | $P(a_1, a_2)$ |

Table 1: Mapping of $DL\text{-}Lite_{\mathcal{R}}$ to FOL

and an ABox $\mathcal{A}$, which is a set of concept and role membership assertions of the forms

$$A(a) \qquad P(a_1, a_2)$$

with $a, a_1, a_2 \in N_{\mathcal{I}}$.
We define the semantics of $DL\text{-}Lite_{\mathcal{R}}$ through a translation to FOL, in the form of the mapping function $\pi$, which is defined in Table 1.[1] The mapping $\pi$ extends naturally to sets of axioms and assertions.

Given a $DL\text{-}Lite_{\mathcal{R}}$ knowledge base $\mathcal{K} = (\mathcal{T}, \mathcal{A})$, the FOL equivalent of $\mathcal{K}$ is the FOL theory $\Phi = \pi(\mathcal{K}) = \pi(\mathcal{T}) \cup \pi(\mathcal{A})$.

*Contextual* $DL\text{-}Lite_{\mathcal{R}}$ is like $DL\text{-}Lite_{\mathcal{R}}$, except that the sets of concept ($N_{\mathcal{C}}$), role ($N_{\mathcal{R}}$), and individual ($N_{\mathcal{I}}$) identifiers are not required to be disjoint. The semantics of a contextual $DL\text{-}Lite_{\mathcal{R}}$ knowledge base $\mathcal{K} = (\mathcal{T}, \mathcal{A})$ is given through the same mapping $\pi(\mathcal{K})$, which yields a contextual FOL theory. Note that contextual $DL\text{-}Lite_{\mathcal{R}}$ is essentially a subset of the QL profile of OWL 2 (Motik, Grau, Horrocks, Wu, Fokoue, & Lutz, 2009a).

## 3. RDF and RDF Schema

We first review the definitions of the RDF syntax and semantics. We then proceed with the embedding of graphs and axiomatization of the entailment regimes into F-Logic, and finally the embedding of extensional RDFS into FOL and $DL\text{-}Lite_{\mathcal{R}}$.

### 3.1 RDF(S) Syntax and Semantics

We proceed with a review of the definitions of the RDF syntax (Klyne & Carroll, 2004) and semantics (Hayes, 2004).

A *vocabulary* $V = \langle \mathcal{C}, \mathcal{PL}, \mathcal{TL} \rangle$ consists of a set $\mathcal{C}$ of RDF URI references (simply referred to as *URIs*), a set $\mathcal{PL}$ of plain literals (i.e., Unicode character strings with an optional language tag), and a set $\mathcal{TL}$ of typed literals (i.e., pairs $(s, u)$ of a Unicode string $s$ and a URI $u$, denoting a *datatype*); see (Klyne & Carroll, 2004, Sections 6.4, 6.5, 6.6) for more details about the specific form of these symbols. Note that $\mathcal{C}$, $\mathcal{PL}$, and $\mathcal{TL}$ are mutually disjoint. The symbols in $V$ are collectively referred to as *names*.

Let $\mathcal{B}$ be a set of blank nodes that is disjoint from the set of names in $V$. Terms are names or blank nodes. A *generalized RDF graph* $S$ is a set of *generalized triples* $\langle s, p, o \rangle$ –

1. Borgida (1996) discusses the relationship between description logics and first-order logic in detail.

*s*ubject, *p*redicate, *o*bject – with $s, p, o \in \mathcal{C} \cup \mathcal{PL} \cup \mathcal{TL} \cup \mathcal{B}$. A *normal RDF graph* $S$ is a set of *normal triples* $\langle s, p, o \rangle$, with $s \in \mathcal{C} \cup \mathcal{B}$, $p \in \mathcal{C}$, and $o \in \mathcal{C} \cup \mathcal{PL} \cup \mathcal{TL} \cup \mathcal{B}$.[2] A *ground triple* is a triple that does not contain blank nodes. A *ground* generalized, respectively normal RDF graph is a set of ground generalized, respectively normal triples. With $bl(\langle s, p, o \rangle) \subseteq \mathcal{B}$ (resp., $bl(S) \subseteq \mathcal{B}$) we denote the set of blank nodes in a triple $\langle s, p, o \rangle$ (resp., graph $S$). In the remainder, whenever speaking about triples or RDF graphs, we mean generalized triples, respectively generalized RDF graphs, unless stated otherwise.

An *interpretation* is a tuple $I = \langle IR, IP, LV, IS, IL, IEXT \rangle$, where $IR$ is a non-empty set, called the domain, $IP$ is a set of properties, $LV \subseteq IR$ is a set of literal values with $\mathcal{PL} \subseteq LV$, $IS$ is a mapping $IS : \mathcal{C} \to IR \cup IP$, $IL$ is a mapping $IL : \mathcal{TL} \to IR$, and $IEXT$ is an extension function $IEXT : IP \to 2^{(IR \times IR)}$.

Given an interpretation $I$, a subset of the blank nodes $\mathcal{B}' \subseteq \mathcal{B}$, and a mapping $A : \mathcal{B}' \to IR$, which is used to interpret blank nodes, for any given term $t$ we define $t^{I,A}$ as:

– if $t \in \mathcal{C}$, then $t^{I,A} = IS(t)$,
– if $t \in \mathcal{PL}$, then $t^{I,A} = t$,
– if $t \in \mathcal{TL}$, then $t^{I,A} = IL(t)$, and
– if $t \in \mathcal{B}'$, then $t^{I,A} = A(t)$.

An interpretation $I$ *satisfies* a triple $\langle s, p, o \rangle$ with respect to a mapping $A : \mathcal{B}' \to IR$, with $bl(\langle s, p, o \rangle) \subseteq \mathcal{B}'$, denoted $(I, A) \models \langle s, p, o \rangle$, if $p^{I,A} \in IP$ and $\langle s^{I,A}, o^{I,A} \rangle \in IEXT(p^{I,A})$. $I$ satisfies a graph $S$ with respect to a mapping $A : bl(S) \to IR$, denoted $(I, A) \models S$, if $(I, A) \models \langle s, p, o \rangle$ for every $\langle s, p, o \rangle \in S$.

An interpretation $I$ satisfies an RDF graph $S$, denoted $I \models S$, if $(I, A) \models S$ for some mapping $A : bl(S) \to IR$; in this case, $I$ is a *model* of $S$. Any interpretation is an *s-interpretation* (simple interpretation).

The notions of *rdf*-, *rdfs*-, and *erdfs*-interpretation are defined through additional conditions on *s*-interpretation. For example, an *s*-interpretation is an *rdf*-interpretation *only if* for every object $k$, $k \in IP$ iff $\langle k, IS(\texttt{rdf:Property}) \rangle \in IEXT(IS(\texttt{rdf:type}))$ and it satisfies the triple $\langle \texttt{rdf:nil}, \texttt{rdf:type}, \texttt{rdf:List} \rangle$. Triples that are required to be satisfied by every $x$-interpretation are called *x-axiomatic triples*, for $x \in \{rdf, rdfs, erdfs\}$ or simply *axiomatic triples* when the entailment regime is clear from the context. The precise definitions of *rdf*-, *rdfs*-, and *erdfs*-interpretation are found in Appendix A.

**Entailment and Satisfiability** Given a vocabulary $V$ and an *entailment regime* $x \in \{s, rdf, rdfs, erdfs\}$, a generalized (resp., normal) RDF graph $S$ *x-entails* a generalized (resp., normal) RDF graph $E$, denoted $S \models_x E$, if every $x$-interpretation of $V$ that is a model of $S$ is also a model of $E$.

Given an entailment regime $x \in \{s, rdf, rdfs, erdfs\}$, a generalized (resp., normal) RDF graph $S$ is *x-satisfiable* if it has a model that is an $x$-interpretation; otherwise it is *x-unsatisfiable*. The following observations can be made about satisfiability for the various entailment regimes; the observations concerning normal RDF graphs are due to Hayes (2004).

**Proposition 2.**
1. *Every generalized and every normal RDF graph is s-satisfiable.*
2. *Every normal RDF graph is rdf-satisfiable.*

2. *Normal RDF graphs* correspond the *RDF graphs* defined by Hayes (2004). In contrast to normal RDF, generalized RDF graphs allow blank nodes and literals in predicate, and literals in subject positions.

*3. There is a generalized RDF graph that is rdf-unsatisfiable.*
*4. There is a normal (and generalized) RDF graph that is rdfs- and erdfs-unsatisfiable.*

### 3.2 Embedding RDF in Logic

We translate a graph to a conjunction of data molecules, where URIs and literals are constant symbols and blank nodes are existentially quantified variables. We axiomatize the entailment regimes using sets of formulas that are independent from the graphs. In the remainder we assume that RDF graphs are finite.

Given a vocabulary $V = \langle \mathcal{C}, \mathcal{PL}, \mathcal{TL} \rangle$, an F-language $\mathcal{L}$ *conforms with* $V$ if it has a signature of the form $\Sigma = \langle \mathcal{C}', \mathcal{P} \rangle$, with $\mathcal{C}' \supseteq \mathcal{C} \cup \mathcal{PL} \cup \mathcal{TL}$.[3]

**Definition 1.** *Let $V$ be a vocabulary, let $S$ be an RDF graph of $V$, let $bl(S) = \{b_1, \ldots, b_n\}$ be the set of blank nodes appearing in $S$, let $\langle s, p, o \rangle$ be a triple in $S$, and let $\mathcal{L}$ be an F-language that conforms with $V$. Then,*

$$\begin{aligned} tr(\langle s,p,o\rangle) &= s[p \twoheadrightarrow o] \text{ and} \\ tr(S) &= \exists\, b_1, \ldots, b_n \big( \bigwedge \{tr(\langle s,p,o\rangle) \mid \langle s,p,o\rangle \in S\} \big) \end{aligned}$$

*are formulas of $\mathcal{L}$.*

The axiomatizations of the entailment regimes are theories $\Psi^x$, with $x \in \{$*s*, *rdf*, *rdfs*, *erdfs*$\}$, which are defined in Appendix B.

If $\phi$ is an F-Logic formula in prenex normal form with only existential quantifiers, then $sk(\phi)$ denotes the *Skolemization* of $\phi$, i.e., every existentially quantified variable is replaced with a globally unique new constant symbol. This extends to theories in the natural way.

**Proposition 3.** *Let $S$ be an RDF graph of a vocabulary $V$ and let $x \in \{s, rdf, rdfs\}$ be an entailment regime. Then, $sk(tr(S)) \cup \Psi^x$ can be equivalently rewritten to a set of F-Logic Horn formulas.*

We have that $\Psi^{erdfs}$ cannot be equivalently rewritten to a set of Horn formulas, because of the use of universal quantification in the antecedents of some of the implications in $\Psi^{erdfs}$.

We now show faithfulness of our embedding.

**Theorem 1.** *Let $S$ and $E$ be RDF graphs of a vocabulary $V$, and let $x \in \{$s, rdf, rdfs, erdfs$\}$ be an entailment regime. Then,*

$$\begin{aligned} S \models_x E \quad &\textit{iff} \quad tr(S) \cup \Psi^x \models_{\mathsf{f}} tr(E) \textit{ and} \\ S \textit{ is } x\textit{-satisfiable} \quad &\textit{iff} \quad tr(S) \cup \Psi^x \textit{ has a model.} \end{aligned}$$

The following corollary follows immediately from Theorem 1 and the classical results about Skolemization (see, e.g., Fitting, 1996).

**Corollary 1.** *Let $S$ and $E$ be RDF graphs of a vocabulary $V$, and let $x \in \{$s, rdf, rdfs, erdfs$\}$ be an entailment regime. Then,*

$$S \models_x E \textit{ if and only if } sk(tr(S)) \cup \Psi^x \models_{\mathsf{f}} tr(E).$$

3. Even though typed literals are pairs in RDF, we treat them simply as constant symbols in our embedding.

Observe that $\Psi^{rdf}$, $\Psi^{rdfs}$, and $\Psi^{erdfs}$ are infinite due to the infinite set of RDF axiomatic triples. However, for checking RDF entailment we need only a finite subset of $\Psi^x$. Given an RDF graph $S$, let $\Psi^x_{-S}$ be obtained from $\Psi^x$ by removing all formulas originating from axiomatic triples involving container membership properties (i.e., `rdf:_1`, `rdf:_2`, ...) not appearing in $S$, with the exception of the axiomatic triples involving `rdf:_1`.

**Proposition 4.** *Let $S$ and $E$ be RDF graphs and let $x \in \{s, rdf, rdfs, erdfs\}$ be an entailment regime. Then,*

$$S \models_x E \text{ if and only if } sk(tr(S)) \cup \Psi^x_{-S\cup E} \models_{\mathsf{f}} tr(E).$$

By Proposition 3 we have that $sk(tr(S)) \cup \Psi^s$, $tr(S)^{sk} \cup \Psi^{rdf}$, and $sk(tr(S)) \cup \Psi^{rdfs}$ are equivalent to sets of Horn formulas. Therefore, Proposition 4 implies that simple, RDF, and RDFS entailment can be computed using reasoners that can compute ground entailment of F-Logic Horn theories, such as FLORA-2 (Yang, Kifer, & Zhao, 2003). Notice that $tr(E)$ can be seen as a boolean conjunctive query (i.e., a yes/no query), where the existentially quantified variables in $tr(E)$ are the non-distinguished variables.

### 3.3 Direct Embedding of Extensional RDFS

We now consider an alternative, direct embedding of the extensional RDFS entailment regime. This embedding, rather than axiomatizing the entailment regime, embeds ontological statements, e.g., `rdfs:subClassOf` statements, directly as formulas.

We first define the notion of *standard use* of the RDF(S) vocabulary, which intuitively corresponds to not using the vocabulary in locations where it can change the semantics of the RDF(S) ontology vocabulary (e.g., ⟨`rdf:type`, `rdfs:subPropertyOf`, $a$⟩).

**Definition 2.** *Let $S$ be an RDF graph. Then, $S$ has only* standard use *of the RDF(S) vocabulary if*

- `rdf:type`, `rdfs:subClassOf`, `rdfs:domain`, `rdfs:range`, *and* `rdfs:subPropertyOf` *do not appear in subject or object positions of any triple in $S$ and*
- `rdfs:ContainerMembershipProperty`, `rdfs:Resource`, `rdfs:Class`, `rdfs:Datatype`, *and* `rdf:Property` *appear only in object positions of* `rdf:type`*-triples in $S$.*

We are now ready to define the direct embedding $tr^{erdfs}$ of the extensional RDFS entailment regime for graphs with only standard use of RDFS vocabulary. While $tr^{erdfs}$ deals with an important part of the RDF(S) vocabulary, the axiomatization of the eRDFS semantics of the remainder of the RDF(S) vocabulary may be found in Appendix B, in the form of the theory $\Psi^{erdfs\text{-}V}$, where $V$ is a vocabulary.

**Definition 3.** *Let $\langle s,p,o\rangle$ be an RDF triple. Then,*

$$\begin{array}{rcl}
tr^{erdfs}(\langle s, \mathtt{type}, \mathtt{Datatype}\rangle) & = & s\!:\!\mathtt{Datatype} \wedge \forall x(x\!:\!s \supset x\!:\!\mathtt{Literal}),\\
tr^{erdfs}(\langle s, \mathtt{type}, \mathtt{ContainerMembership\text{-}} & = & s\!:\!\mathtt{ContainerMembershipProperty}\\
\mathtt{Property}\rangle) & & \wedge\forall x,y(x[s \twoheadrightarrow y] \supset x[\mathtt{member} \twoheadrightarrow y]),\\
tr^{erdfs}(\langle s, \mathtt{type}, o\rangle) & = & s\!:\!o,\\
tr^{erdfs}(\langle s, \mathtt{subClassOf}, o\rangle) & = & \forall x(x\!:\!s \supset x\!:\!o),\\
tr^{erdfs}(\langle s, \mathtt{subPropertyOf}, o\rangle) & = & \forall x,y(x[s \twoheadrightarrow y] \supset x[o \twoheadrightarrow y]),\\
tr^{erdfs}(\langle s, \mathtt{domain}, o\rangle) & = & \forall x,y(x[s \twoheadrightarrow y] \supset x\!:\!o),\\
tr^{erdfs}(\langle s, \mathtt{range}, o\rangle) & = & \forall x,y(x[s \twoheadrightarrow y] \supset y\!:\!o),\ \textit{and}\\
tr^{erdfs}(\langle s, p, o\rangle) & = & s[p \twoheadrightarrow o],\ \textit{otherwise.}
\end{array}$$

*Let $S$ be an RDF graph and let $bl(S) = \{b_1, \dots, b_n\}$ be the set of blank nodes in $S$. Then,*

$$tr^{erdfs}(S) \;=\; \{\exists\, b_1,\dots,b_n(\textstyle\bigwedge\{tr^{erdfs}(\langle s,p,o\rangle) \mid \langle s,p,o\rangle \in S\})\}$$

We say that a term $t$ occurs in a *property position* if it occurs as the predicate of a triple, as the subject or object of an `rdfs:subPropertyOf` triple, as the subject of an `rdfs:domain` or `rdfs:range` triple, or the graph contains $\langle t, \mathtt{rdf{:}type}, \mathtt{rdf{:}Property}\rangle$ or $\langle t, \mathtt{rdf{:}type}, \mathtt{ContainerMembershipProperty}\rangle$. A term $t$ occurs in a *class position* if it occurs as the subject or object of an `rdfs:subClassOf` triple, as the object of an `rdfs:domain`, `rdfs:range`, or `rdf:type` triple, as the subject of a triple $\langle t, \mathtt{rdf{:}type}, \mathtt{rdfs{:}Class}\rangle$, or as the subject of a triple $\langle t, \mathtt{rdf{:}type}, \mathtt{rdfs{:}Datatype}\rangle$.

Let $S$ be an RDF graph with only standard use of the RDF(S) vocabulary. The *property* (resp., *class*) *vocabulary* of $S$ consists of all the names appearing in property (resp., class) positions in $S$ or the RDF(S) axiomatic triples with only standard use of the RDF(S) vocabulary.

Given two RDF graphs $S$ and $E$ with only standard use of the RDF(S) vocabulary, we write $E \trianglelefteq S$ if the property, resp. class vocabularies of $E$ are subsets of the property, resp. class vocabularies of $S$, there are no blank nodes in class or property positions in $E$,[4] and `rdfs:Resource`, `rdfs:Class`, and `rdf:Property` do not appear in $E$.

**Theorem 2.** *Let $S$ and $E$ be RDF graphs with only standard use of the RDFS vocabulary such that $E \trianglelefteq S$. Then,*

$$S \models_{erdfs} E \;\; \textit{iff} \;\; tr^{erdfs}(S) \cup \Psi^{erdfs\text{-}V} \models_{\mathsf{f}} tr^{erdfs}(E)$$

We define $\Psi^{erdfs\text{-}V}_{-S}$ analogously to $\Psi^{erdfs}_{-S}$, i.e., it does not contain statements concerning container membership properties not appearing in the graph $S$, with the exception of `rdf:_1`. The following proposition follows from an argument analogous to the proof of Property 4.

**Proposition 5.** *Let $S$ and $E$ be RDF graphs with only standard use of the RDFS vocabulary such that $E \trianglelefteq S$. Then,*

$$S \models_{erdfs} E \;\; \textit{iff} \;\; sk(tr^{erdfs}(S)) \cup \Psi^{erdfs\text{-}V}_{-S\cup E} \models_{\mathsf{f}} tr^{erdfs}(E).$$

4. This restriction on the use of blank nodes in the entailed graph was not mentioned in the extended abstract of this paper (de Bruijn & Heymans, 2007). This was an error.

We have that whenever $E$ does not contain the terms `rdfs:subClassOf`, `rdfs:subPropertyOf`, `rdfs:domain`, and `rdfs:range`, $tr^{erdfs}(E)$ is a conjunction of atomic molecules prefixed by an existential quantifiers (i.e., a conjunctive query).

We have that $sk(tr^{erdfs}(S)) \cup \Psi^{erdfs\text{-}V}_{-S \cup E}$ is a finite set of Horn formulas. Therefore, if the graphs satisfy the mentioned conditions, query answering techniques used in F-Logic reasoners such as FLORA-2 (Yang et al., 2003) can be used for checking extensional RDFS entailment.

### 3.4 Embedding Extensional RDFS into First-Order Logic

We now consider an embedding of extensional RDFS entailment into first-order logic (FOL), based on the direct embedding of extensional RDFS in F-Logic defined above.

We say that an F-Logic theory $\Phi$ is *translatable* to contextual FOL if $\Phi$ does not contain unary or binary predicates and for every molecule of the form $t_1[t_2 \twoheadrightarrow t_3]$ or $t_1 : t_2$ holds that $t_2$ is a constant symbol. $FO(\Phi)$ is the contextual FOL theory obtained from $\Phi$ by replacing:

- every data molecule $t_1[t_2 \twoheadrightarrow t_3]$ with the atomic formula $t_2(t_1, t_3)$ and
- every is-a molecule $t_1 : t_2$ with the atomic formula $t_2(t_1)$.

The following proposition follows immediately from an earlier result (de Bruijn & Heymans, 2008, Theorem 3.2).

**Proposition 6.** *Let $\Phi$, respectively $\phi$, be an equality-free F-Logic theory, respectively formula, that is translatable to contextual FOL. Then,*

$$\Phi \models_{\mathsf{f}} \phi \textit{ iff } FO(\Phi) \models_c FO(\phi).$$

We say that an RDF graph $S$ is a *non-higher-order* graph if $S$ does not contain blank nodes in class or property positions, and has only standard use of the RDFS vocabulary. Observe that if $S$ is a non-higher-order RDF graph, then $tr^{erdfs}(S) \cup \Psi^{erdfs\text{-}V}$ is translatable to contextual FOL. Notice also that every ground RDF graph that has only standard use of the RDFS vocabulary is a non-higher-order RDF graph.

**Theorem 3.** *Let $S$ and $E$ be non-higher-order RDF graphs such that $E \trianglelefteq S$. Then,*

$$S \models_{erdfs} E \textit{ iff } FO(tr^{erdfs}(S) \cup \Psi^{erdfs\text{-}V}) \models_c FO(tr^{erdfs}(E)).$$

*Proof.* Follows immediately from Theorem 2, the fact that $FO(tr^{erdfs}(S))$ and $FO(tr^{erdfs}(E))$ do not contain the equality symbol, and Proposition 6. □

Concerning the relationship with *DL-Lite*$_{\mathcal{R}}$, we make the following observation.

**Proposition 7.** *Let $S$ be a ground non-higher-order graph.*[5] *Then, $FO(tr^{erdfs}(S) \cup \Psi^{erdfs\text{-}V})$ can be equivalently rewritten to the FOL equivalent $\Phi = \pi(\mathcal{K})$ of a contextual DL-Lite$_{\mathcal{R}}$ knowledge base $\mathcal{K}$.*

Analogous to Proposition 5, one may discard the axiomatic triples concerning container membership properties that are not used, and thus one only needs to reason with a finite knowledge base.

5. Note that, when considering a variant of *DL-Lite*$_{\mathcal{R}}$ that allows existentially quantified variables in the ABox – also allowed in OWL DL – this restriction could be relaxed to $S$ being a non-ground non-higher-order RDF graph.

## 4. Extensions with Datatypes

The entailment regimes we dealt with in the previous section do not consider many of the useful datatypes (e.g., strings, integers). In fact, `rdf:XMLLiteral` is the only datatype that was considered. The RDF semantics specification (Hayes, 2004) defines the notion of $D$ entailment (datatype entailment), which extends RDFS entailment with support for datatypes. Ter Horst (2005) defines the notion of $D*$ entailment, which is also an extension of RDFS entailment, but semantically weaker than $D$ entailment. We first review $D*$ entailment, after which we review $D$ entailment. Both semantics were originally defined as extensions of RDFS entailment. However, one might extend any of the entailment regimes we considered with datatype support. Therefore, we consider extensions of simple, RDF, RDFS, and extensional RDFS entailment with both kinds of datatype semantics. We first review the datatype semantics, after which we present embeddings of both semantics into F-Logic. Finally, we discuss a notion of normalization that may be used to remove equality statements from the embeddings to speed up processing.

### 4.1 Extension of the RDF Entailment Regimes with Datatypes

Datatypes define sets of concrete data values (e.g., strings and integers), along with their lexical representations. A *datatype* is a tuple $d = \langle L^d, V^d, L2V^d \rangle$ consisting of

- a *lexical space* $L^d$, which is a set of character strings (e.g., "0", "1", "01", ..., in the case of an integer datatype),
- a *value space* $V^d$, which is a set of values (e.g., the numbers 0, 1, 2, ..., in the case of an integer datatype), and
- a *lexical-to-value mapping* $L2V^d$, which is a mapping from the lexical space to the value space (e.g., $\{$"0" $\mapsto 0$, "1" $\mapsto 1$, "01" $\mapsto 1, \ldots\}$, for an integer datatype).

A *simple datatype map* $D$ is a partial mapping from URIs to datatypes. A simple datatype map $D$ is a *datatype map* if $D(\texttt{rdf:XMLLiteral}) = xml$ where $xml$ is the built-in XML literal datatype as defined in the RDF specification (Klyne & Carroll, 2004). With $dom(D)$ and $ran(D)$ we denote the domain and range of $D$, respectively.

Given a simple datatype map $D$, we call a typed literal $(s, u) \in \mathcal{TL}$ *well-typed* if $u \in dom(D)$ and $s \in L^{D(u)}$; $(s, u)$ is *ill-typed* if $u \in dom(D)$ and $s \notin L^{D(u)}$.

We now review the notions of $D*$ and $D$ entailment. Similar to the previous section, the definitions of $D*$- and $D$-interpretations can be found in Appendix A.

***D** entailment** Given a simple datatype map $D$, an RDF graph $S$ *s-$D*$ entails* an RDF graph $E$, denoted $S \models_{s\text{-}D*} E$, if every $s$-$D*$-interpretation that is a model of $S$ is a model of $E$.

Given a datatype map $D$, an RDF graph $S$ *x-$D*$-entails* an RDF graph $E$, denoted $S \models_{x\text{-}D*} E$, if every $x$-$D*$-interpretation that is a model of $S$ is a model of $E$, for $x \in \{rdf, rdfs, erdfs\}$.

Notice that if $dom(D) = \{\texttt{rdf:XMLLiteral}\}$ and $x \in \{rdf, rdfs, erdfs\}$, then $x$-$D*$-entailment corresponds to $x$-entailment, with the exception that when considering $rdf$-$D*$-entailment, the triple $\langle \texttt{rdf:XMLLiteral}, \texttt{rdf:type}, \texttt{rdfs:Datatype} \rangle$ is additionally entailed. In addition, if $dom(D) = \emptyset$, then $s$-$D*$-entailment corresponds to $s$-entailment.

The following example shows how equality may be introduced by the $D^*$ semantics.

**Example 2.** *Consider a datatype map that contains the XML schema* `string` *datatype (Peterson, Gao, Malhotra, Sperberg-McQueen, & Thompson, 2009). Certain equalities hold between plain literals without language tags and typed literals of this datatype, because the set of plain literals without language tags corresponds to the value space of the* `string` *datatype. So, equalities such as "a" = ("a",* `string`*) and "xxx" = ("xxx",* `string`*) necessarily hold. Similar for equalities between datatypes. For example, if the datatype map contains both* `integer` *and* `decimal`*, then further equalities such as ("1",* `integer`*) = ("1",* `decimal`*) and ("1",* `decimal`*) = ("1.0",* `decimal`*) necessarily hold.*

***D* entailment** Given a simple datatype map $D$, an RDF graph $S$ *s-D-entails* an RDF graph $E$, denoted $S \models_{s\text{-}D} E$, if every $s$-$D$-interpretation which is a model of $S$ is a model of $E$.

Given a datatype map $D$, an RDF graph $S$ *x-D-entails* an RDF graph $E$, denoted $S \models_{x\text{-}D} E$, if every $x$-$D$-interpretation which is a model of $S$ is a model of $E$, for $x \in \{rdf, rdfs, erdfs\}$. An RDF graph $S$ *x-D-entails* an RDF graph $E$, denoted $S \models_{x\text{-}D} E$, if every $x$-$D$-interpretation which is a model of $S$ is also a model of $E$.

There are two main differences between $D^*$ entailment and $D$ entailment: (i) $D$ entailment allows for easy extension towards languages which can express equality between URIs denoting datatypes; whenever two URIs denote the same datatype, typed literals with these two URIs as datatypes are interpreted in the same way (see Example 3); and (ii) $D$ entailment directly links the class extension of a datatype with the value space of this datatype. The latter complicates the evaluation of entailment somewhat, and was likely the main motivation for the introduction of $D^*$ entailment. The complication becomes clear when declaring blank nodes as members of specific datatypes, as illustrated in Example 4.

**Example 3.** *Consider an extension of D entailment with equality by imposing the following condition on interpretations:*

> *(+) An interpretation I satisfies a triple $\langle x, \texttt{owl:sameAs}, y\rangle$ with respect to a blank node assignment A iff $x^{I,A} = y^{I,A}$.*

*Now consider a datatype map $D = \{bool \mapsto boolean\}$, where boolean is defined as follows:*

- $L^{boolean} = \{\text{"1"}, \text{"0"}, \text{"t"}, \text{"f"}\}$,
- $V^{boolean} = \{true, false\}$, *and*
- $L2V^{boolean} = \{\text{"1"} \mapsto true, \text{"0"} \mapsto false, \text{"t"} \mapsto true, \text{"f"} \mapsto false\}$,

*and an RDF graph $S = \{\langle myBool, \texttt{owl:sameAs}, bool\rangle, \langle a, b, (\text{"1"}, myBool)\rangle\}$. In D-interpretations, typed literals of which the datatype URIs are interpreted the same are interpreted the same as well. Therefore, under D entailment extended with condition (+) the triple $\langle a, b, (\text{"t"}, myBool)\rangle$ can be derived from S: ("1", myBool) and ("t", myBool) are both interpreted as $L2V^{boolean}(\text{"1"}) = L2V^{boolean}(\text{"t"}) = true$; hence, ("1", myBool) and ("t", myBool) are interpreted in the same way in every interpretation. Similarly, it can be shown that the triples $\langle a, b, (\text{"1"}, bool)\rangle$ and $\langle a, b, (\text{"t"}, bool)\rangle$ are entailed by S.*

*None of these derivations is valid when considering $D^*$ entailment extended with condition (+). In fact, because myBool is not in the domain of D, ("1", myBool) is interpreted as an arbitrary (abstract) symbol; it is treated in the same way as a URI.*

**Example 4.** *Consider a datatype map $D$ that includes the XML schema datatypes* `string` *and* `integer` *(Peterson et al., 2009), which have disjoint value spaces. Consider also the graph* $S = \{\langle _ : x, \texttt{rdf:type}, \textit{string}\rangle, \langle _ : x, \texttt{rdf:type}, \textit{integer}\rangle\}$. *In an rdfs-$D^*$-interpretation $I$ the class extensions of* `string` *and* `integer` *are not necessarily the same as the value spaces of the respective datatypes. Therefore, there may be an object $k \in IR$ that is neither an integer nor a string, but which is in the class extensions of both* `string` *and* `integer`. *Consequently, there is an rdfs-$D^*$-interpretation that is a model of $S$ and $S$ is rdfs-$D^*$-satisfiable.*

*In an rdfs-D-interpretation the class extensions of* `string` *and* `integer` *are necessarily the same as the value spaces of the respective datatypes. Since these value spaces are disjoint, there can be no object that is both in the class extension of* `string` *and in the class extension of* `integer`. *Therefore, $S$ is not rdfs-D-satisfiable.*

### 4.2 Embeddings of Datatypes in Logic

Given a datatype map $D$, we use a set of formulas $\Psi^y \subseteq \mathcal{L}$, defined in Appendix B, to axiomatize the semantics of an entailment regime $y \in \{x\text{-}D^*, x\text{-}D\}$, with $x \in \{s, rdf, rdfs, erdfs\}$.

Analogous to Proposition 3, we have:

**Proposition 8.** *Let $S$ be an RDF graph of a vocabulary $V$. Then, $sk(tr(S)) \cup \Psi^y$, with $y \in \{s\text{-}D^*, rdf\text{-}D^*, rdfs\text{-}D^*, s\text{-}D, rdf\text{-}D, rdfs\text{-}D\}$, can be equivalently rewritten to a set of F-Logic Horn formulas.*

We first show faithfulness of our embedding of $D^*$ entailment.

**Theorem 4.** *Let $S$ and $E$ be RDF graphs of a vocabulary $V$, let $D$ be a datatype map, and let $x \in \{s, rdf, rdfs, erdfs\}$ be an entailment regime. Then,*

$$S \models_{x\text{-}D^*} E \text{ if and only if } tr(S) \cup \Psi^{x\text{-}D^*} \models_{\mathsf{f}} tr(E) \text{ and}$$

$$S \text{ is } x\text{-}D^*\text{-satisfiable iff } tr(S) \cup \Psi^{x\text{-}D^*} \text{ has a model.}$$

We now turn to $x$-$D$-entailment. It turns out that when considering datatype maps with arbitrary datatypes, one needs to reason by case (see Proposition 14 in Section 5), which complicates matters. We therefore restrict ourselves to *definite* datatypes, which do not bring this complication. An example of a definite datatype map is one that includes only the set of datatypes in the OWL 2 EL and QL profiles (Motik et al., 2009a).

**Definition 4.** *A datatype map $D$ is* definite *if*

- *the value space of every datatype $d \in ran(D)$ is infinite,*
- *for any $n \geq 1$ distinct datatypes $d_1, \ldots, d_n \in ran(D)$ holds that either (a) the value spaces are disjoint, i.e., $V^{d_i} \cap V^{d_j} = \emptyset$ $(1 \leq i < j \leq n)$ or (b) their intersection is infinite, i.e., $V^{d_1} \cap \cdots \cap V^{d_n}$ is an infinite set, and*

- *for no two datatypes $d_1, d_2 \in ran(D)$ holds that $d_1 \in V^{d_2}$.*

**Theorem 5.** *Let $S$ and $E$ be RDF graphs of a vocabulary $V$, let $D$ be a definite datatype map, and let $x \in \{s, rdf, rdfs, erdfs\}$ be an entailment regime. Then,*

$$S \models_{x\text{-}D} E \textit{ if and only if } tr(S) \cup \Psi^{x\text{-}D} \models_{\mathsf{f}} tr(E) \textit{ and}$$

$$S \textit{ is } x\text{-}D\textit{-satisfiable iff } tr(S) \cup \Psi^{x\text{-}D} \textit{ has a model.}$$

### 4.3 Normalization of Datatypes

The set of equality statements in the axiomatizations $\Psi^{x\text{-}D^*}$ and $\Psi^{x\text{-}D}$ is potentially large and, in general, polynomial in the size of the vocabulary $V$. In addition, it requires equality reasoning, which tends to deteriorate the performance of a reasoner. We discuss how to normalize the embedding of a graph in F-Logic, thereby removing the need for expressing equality.

Given a vocabulary $V$, we assume a strict (e.g., lexicographical) order $<$ on the set of literals $\mathcal{PL} \cup \mathcal{TL}$. For a given datatype map $D$, we define $V^D = \bigcup_{u \in dom(D)} V^{D(u)}$, i.e., the values in $D$. For each $v \in V^D$, we define the literals that represent the value $v$ as: $\overline{v} = \{(s,u) \in \mathcal{TL} \mid L2V^{D(u)}(s) = v\} \cup \{l \in \mathcal{PL} \mid l = v\}$. The *representation* of $\overline{v}$, denoted $r(\overline{v})$, is the least element in $\overline{v}$ according to the order $<$.

Given a set of formulas $\Phi \subseteq \mathcal{L}$ such that $\mathcal{L}$ conforms with $V$, the *datatype normalization* of $\Phi$, denoted $(\Phi)^{\mathsf{n}}$, is obtained from $\Phi$ by replacing every plain literal $l \in \mathcal{PL}$ with $r(\bar{l})$ and replacing every well-typed literal $(s,u) \in \mathcal{TL}$ with $r(\overline{L2V^{D(u)}(s)})$.

Observe that the only equality statements in the normalizations $(tr(S) \cup \Psi^{x\text{-}D})^{\mathsf{n}}$ and $(tr(S) \cup \Psi^{x\text{-}D^*})^{\mathsf{n}}$ are trivial statements of the form $t = t$, where $t$ is a literal. Therefore, these statements may be discarded.

The following proposition follows straightforwardly from the shape of the axiomatizations $\Psi^y$ and the definition of normalization.

**Proposition 9.** *Let $S$ and $E$ be RDF graphs of a vocabulary $V$, let $D$ be datatype map $D$, and let $y \in \{$s-D*, rdf-D*, rdfs-D*, erdfs-D*, s-D, rdf-D, rdfs-D, erdfs-D$\}$. Then,*

$$tr(S) \cup \Psi^y \models_{\mathsf{f}} tr(E) \textit{ iff } (tr(S) \cup \Psi^y)^{\mathsf{n}} \models_{\mathsf{f}} (tr(E))^{\mathsf{n}}$$

## 5. Complexity

In this section we review the complexity of the various RDF entailment relations and present several novel results, exploiting the embeddings presented in Sections 3 and 4.

The complexity of non-ground simple entailment and RDFS entailment, and upper bounds for ground entailment are known from the literature, and analogous results for RDF entailment follow immediately. Recall that, although the set of axiomatic triples is infinite, only a finite subset, linear in the size of the graphs, needs to be taken into account when checking entailment (cf. Proposition 4).

**Proposition 10** (Gutierrez et al., 2004, 2010; ter Horst, 2005; de Bruijn et al., 2005)**.** *The decision problems $S \models_s E$, $S \models_{rdf} E$, $S \models_{rdfs} E$, and $S \models_{rdfs\text{-}D^*} E$, given RDF graphs $S$* *and $E$, are* NP-complete *in the combined size of $S$ and $E$, and polynomial in the size of $S$. If $E$ is ground, then the respective problems are in* PTime.
*In addition, the problems $S \models_{erdfs} E$ and $S \models_{rdfs\text{-}D} E$ are* NP-hard.

The membership proofs by Gutierrez et al. (2004, 2010), ter Horst (2005), and de Bruijn et al. (2005) rely on the fact that the set of all (relevant) entailed triples of a given graph can be computed in polynomial time using the RDFS entailment rules (ter Horst, 2005); the problem can then be reduced to subgraph homomorphism. From Corollary 1 and the fact that the problem of checking ground entailment in Datalog (Dantsin, Eiter, Gottlob, & Voronkov, 2001) is polynomial in the size of the data (i.e., $tr(S)$) we obtain a novel argument for membership.

NP-hardness of non-ground entailment has been shown through a reduction from a known NP-hard problem (ter Horst, 2005).

From the embedding in F-Logic (Theorem 1), we obtain the following upper bound for the complexity of simple and RDF entailment.

**Proposition 11.** *Let $S$ and $E$ be RDF graphs. If $E$ is fixed, the problems $S \models_s E$ and $S \models_{rdf} E$ are decidable in* LogSpace *in the size of $S$. The problems $S \models_s E$ and $S \models_{rdf} E$ are decidable in* LogSpace *in the combined size of the graphs if $E$ is ground.*

*Proof Sketch.* It is easy to see that the only fact that could potentially be recursively derived from $\Psi^{rdf}$ is $\texttt{rdf:type}[\texttt{rdf:type} \twoheadrightarrow \texttt{rdf:Property}]$; however, $\texttt{rdf:type}[\texttt{rdf:type} \twoheadrightarrow \texttt{rdf:Property}] \in \Psi^{rdf}$. Thus, $sk(tr(S))$ and $sk(tr(S)) \cup \Psi^{rdf}$ may be treated as nonrecursive Datalog programs.

The proposition then follows straightforwardly from Corollary 1 and the fact that ground entailment in nonrecursive Datalog is in LogSpace in the size of the data (Abiteboul, Hull, & Vianu, 1995), with the data being the input RDF graphs. □

It turns out that we cannot obtain a LogSpace upper bound for RDFS entailment. In fact, it turns out that ground *rdfs*-, and hence ground *rdfs-D**- and *rdfs-D*-entailment, is PTime-hard.

**Proposition 12.** *There exist ground RDF graphs $S$ and $E$ such that the decision problems $S \models_{rdfs} E$, $S \models_{rdfs\text{-}D^*} E$, and $S \models_{rdfs\text{-}D} E$ are* PTime-hard.

*Proof.* We proceed by reduction from the PTime-hard problem path system accessibility (Jones & Laaser, 1974; Gary & Johnson, 1979), which is defined as:
Instance: A set $X$ of nodes, subsets $S, T \subseteq X$ of source and terminal nodes, and a relation $R \subseteq X \times X \times X$.
Question: A node $x \in X$ is *accessible* if $x \in S$ or if there exist accessible nodes $y, z \in X$ such that $\langle x, y, z \rangle \in R$. Is there an accessible terminal node $t \in T$?

In the remainder sp is short for rdfs:subPropertyOf.

We now encode this problem into RDFS. The graph $G$ is constructed as follows:

- for every source node $x \in S$ include the triple $\langle x, \texttt{sp}, \texttt{sp} \rangle$,
- for every terminal node $x \in T$ include the triple $\langle a, \texttt{sp}, x \rangle$, and
- for every tuple $\langle x, y, z \rangle \in R$ include the triple $\langle x, y, z \rangle$.

We show that a node $t \in X$ is accessible iff $G \models_{rdfs} \langle t, \mathtt{sp}, \mathtt{sp}\rangle$. It follows that there exists an accessible node iff $G \models_{rdfs} \langle a, \mathtt{sp}, \mathtt{sp}\rangle$.

($\Rightarrow$) We proceed by induction. Base case: if $t \in S$ then $\langle t, \mathtt{sp}, \mathtt{sp}\rangle \in G$, so clearly $G \models_{rdfs} \langle t, \mathtt{sp}, \mathtt{sp}\rangle$.

Induction step: consider $\langle t, y, z\rangle \in R$ such that $y, z$ are accessible. We have that $\langle t, y, z\rangle$ is included in $G$ and $G \models_{rdfs} \langle y, \mathtt{sp}, \mathtt{sp}\rangle$ and $G \models_{rdfs} \langle z, \mathtt{sp}, \mathtt{sp}\rangle$, since $y$ and $z$ are accessible. Condition 10 in Table 5 implies that $G \models_{rdfs} \langle t, \mathtt{sp}, z\rangle$. By transitivity of sp (condition 9 in Table 5) we can subsequently conclude that $G \models_{rdfs} \langle t, \mathtt{sp}, \mathtt{sp}\rangle$.

($\Leftarrow$) Assume, on the contrary, that $t \in X$ is not accessible. It is then straightforward to construct an *rdfs*-interpretation $I$ such that $I \models_{rdfs} G$ and $I \not\models_{rdfs} \langle t, \mathtt{sp}, \mathtt{sp}\rangle$, a contradiction. □

Using the correspondence of Proposition 7, the results on the complexity of reasoning in $DL\text{-}Lite_{\mathcal{R}}$ (Calvanese et al., 2007), and the classical results on skolemization (Fitting, 1996) we obtain the following result for extensional RDFS entailment. Recall the notion of standard use of the RDFS vocabulary from Definition 2.

**Proposition 13.** *Let $S$ and $E$ be RDF graphs with only standard use of the RDFS vocabulary such that $E \trianglelefteq S$. Then, the problem of deciding $S \models_{erdfs} E$ is NP-complete, and NLogSpace-complete if $E$ is ground.*

*Proof.* Assume that $E$ is ground. We first demonstrate membership.

We have that $FO(sk(tr_{erdfs}(S)) \cup \Psi^{erdfs\text{-}V})$ is a theory of contextual FOL that is equivalent to a contextual $DL\text{-}Lite_{\mathcal{R}}$ knowledge base (by Proposition 7). If $E$ is ground, then, as a straightforward consequence from Theorems 2 and 3,

$$S \models_{erdfs} E \text{ iff } FO(sk(tr^{erdfs}(S)) \cup \Psi^{erdfs\text{-}V}_{-S\cup E}) \models_c FO(tr^{erdfs}(E)).$$

A contextual $DL\text{-}Lite_{\mathcal{R}}$ theory $\Phi^c$ (resp., formula $\phi^c$) can be straightforwardly rewritten to a corresponding classical $DL\text{-}Lite_{\mathcal{R}}$ theory $\Phi$ (resp., formula $\phi$) such that

$$\Phi^c \models_c \phi^c \text{ iff } \Phi \models \phi.$$

Since this transformation is linear in the size of the knowledge base, the complexity of deciding satisfiability and entailment of contextual $DL\text{-}Lite_{\mathcal{R}}$ knowledge bases is the same as that of $DL\text{-}Lite_{\mathcal{R}}$ knowledge bases, namely NLogSpace (Calvanese et al., 2007).

Hardness is shown by reduction from a known NLogSpace-hard problem: Graph reachability (Papadimitriou, 1994) can be encoded using subclass statements: edges in the graph are represented in the RDF graph $S$ by `rdfs:subClassOf`-triples and $t$ is reachable from $s$ iff $S \models_{erdfs} \{\langle s, \mathtt{rdfs{:}subClassOf}, t\rangle\}$.

This result immediately leads to the following NP algorithm for deciding $S \models_{erdfs} E$, in case $E$ is not ground:

1. Guess a mapping $\theta$ from blank nodes in $E$ to ground terms in $FO(sk(tr^{erdfs}(S)) \cup \Psi^{erdfs\text{-}V}_{-S\cup E})$.
2. Check whether $FO(sk(tr^{erdfs}(S)) \cup \Psi^{erdfs\text{-}V}_{-S\cup E}) \models_c FO(tr^{erdfs}(E)\theta)$.

This algorithm is clearly sound and complete, since the theory $FO(sk(tr^{erdfs}(S)) \cup \Psi^{erdfs\text{-}V}_{-S\cup E})$ is universal.

NP-hardness follows from NP-hardness of simple entailment (Proposition 10), which is straightforwardly encoded into extensional RDFS entailment. □

For $x$-$D$-entailment with arbitrary datatype maps we obtain the following novel lower bound.

**Proposition 14.** *There are RDF graphs $S$ and $E$ and a datatype map $D$ such that deciding $S \models_{s\text{-}D} E$ is coNP-hard in the size of $S$.*

*Proof.* We proceed by reduction from the complement of GRAPH $k$-COLORABILITY for $k \geq 3$, i.e., the nonexistence of a $k$-coloring. This problem is coNP-complete (Gary & Johnson, 1979):

Instance: A graph $G = \langle V, E \rangle$ and a positive integer $k \leq |V|$ such that $k \geq 3$.

Question: A $k$-coloring is an assignment from nodes to colors $f : V \to \{1, 2, \ldots, k\}$ such that no two adjacent nodes share the same color, i.e., if $\langle u, v \rangle \in E$, then $f(u) \neq f(v)$. Is it the case that there is no $k$-coloring?

Let $D$ be a datatype map that includes `rdf:XMLLiteral` and that maps a URI `d` to a datatype $D(\mathtt{d})$ with an ordered value space of cardinality $k$, let $S$ be the smallest RDF graph such that:

- for every $v \in V$, $\langle v, \mathtt{rdf{:}type}, \mathtt{d} \rangle \in S$ and
- for every $\langle u, v \rangle \in E$, $\langle u, \mathtt{R}, v \rangle \in S$,

where `R` is a URI, and let $H = \{\langle _{:}x, \mathtt{R}, _{:}x \rangle\}$, where $_{:}x$ is a blank node.

We now show that $G$ does not have a $k$-coloring if and only if $S \models_{s\text{-}D} H$.

($\Rightarrow$) Assume, on the contrary, that $S \not\models_{s\text{-}D} H$, which means there is an $s$-$D$-interpretation $I$ such that $I \models S$ and $I \not\models H$. Therefore, (*) there is no $s \in IR$ such that $\langle s, s \rangle \in IEXT(IS(\mathtt{R}))$. Consider any $\langle u, v \rangle \in E$; by (*) we have that $IS(u) \neq IS(v)$. Since $\langle u, \mathtt{rdf{:}type}, \mathtt{d} \rangle$, $\langle v, \mathtt{rdf{:}type}, \mathtt{d} \rangle \in S$, $IS(u), IS(v) \in D(d)$, by condition 20 in Table 8. Now let $f(v) = IS(v)$ for every $v \in V$. We have that $f$ is a $k$-coloring, a contradiction.

($\Leftarrow$) Analogously, if there exists a $k$-coloring, one can construct an $s$-$D$-interpretation that is a model of $S$, but not of $H$. □

A *polynomial* (resp., *logspace*) *datatype map* $D$ is a datatype map for which holds that deciding well-typedness of literals and deciding $L2V^{D(u)}(s) = L2V^{D(u')}(s')$ and $l = L2V^{D(u)}(s)$, where $l$ is a plain literal and $(s, u), (s', u')$ are well-typed literals, can be done in PTime (resp., LogSpace).

Considering definite datatype maps, we obtain the following lower bound from Theorem 5 and the data complexity of Datalog, exploiting Skolemization, analogous to Corollary 1, and exploiting the fact that we need to take into account only a subset of the RDF(S) axiomatic triples, analogous to Proposition 4.

**Proposition 15.** *Let $D$ be a definite polynomial datatype map. Then, the decision problems $S \models_{s\text{-}D} E$, $S \models_{rdf\text{-}D} E$, and $S \models_{rdfs\text{-}D} E$ are NP-complete in the combined size of $S$ and $E$, and polynomial in the size of $S$. If $E$ is ground, then the respective problems are in PTime.*

It turns out that, analogous to the case without datatypes, we can further refine the upper bounds of simple- and *rdf*-entailment.

**Lemma 1.** *Let $\Phi$ be a theory and let $D$ be a logspace datatype map. Then, $(\Phi)^{\mathsf{n}}$ can be computed in LogSpace.*

| Entailment | Restrictions on $S$ | Restrictions on $E$ | Complexity |
|---|---|---|---|
| $\models_s$,$\models_{rdf}$,$\models_{rdfs}$ | none | none | NP-complete |
| $\models_s$,$\models_{rdf}$ | none | ground | LogSpace |
| $\models_{rdfs}$ | none | ground | P-complete |
| $\models_{erdfs}$ | none | none | NP-hard |
| $\models_{erdfs}$ | stand. RDFS | stand. RDFS | NP-complete |
| $\models_{erdfs}$ | stand. RDFS | stand. RDFS, ground | NLogSpace-complete |

Table 2: Complexity of Entailment $S \models_x E$ in RDF, measured in the combined size of $S$ and $E$

| Entailment | $\emptyset$ | $D^*$ | definite $D$ | $D$ |
|---|---|---|---|---|
| x=$s$ | LogSpace | LogSpace | LogSpace | coNP-hard |
| x=$rdf$ | LogSpace | LogSpace | LogSpace | coNP-hard |
| x=$rdfs$ | P-complete | P-complete | P-complete | coNP-hard |

Table 3: Complexity of Entailment $S \models_{x\text{-}D} E$ and $S \models_{x\text{-}D^*} E$, measured in the size of $S$

*Proof.* With $\mathcal{WL}$ we denote the set of plain and well-typed literals, and with $<$ the lexicographical ordering over $\mathcal{WL}$. If $l$ is a plain literal, we define $v^l = l$; if $(s, u)$ is a well-typed literal, $v^{(s,u)} = L2V^{D(u)(s)}$. The following algorithm returns the representation of a literal $l \in \mathcal{WL}$ in LogSpace: iterate over all literals $l' < l$, until the least literal $l''$ such that $v^{l''} = v^l$ is found; observe that deciding $l' < l$ and deciding $v^{l''} = v^l$ can be done in LogSpace. □

From the lemma we obtain the following upper bound, by considerations analogous to Proposition 11 and the fact that the axioms in $\Psi^D \backslash \Psi^{rdf}$ do not introduce recursion.

**Proposition 16.** *Let $S$ and $E$ be RDF graphs and let $D$ be a logspace datatype map. Then, the problems $S \models_{s\text{-}D^*} E$ and $S \models_{rdf\text{-}D^*} E$ are decidable in LogSpace in the size of $S$, and in the combined size of the graphs if $E$ is ground.*

*Furthermore, if $D$ is definite, the problems $S \models_{s\text{-}D} E$ and $S \models_{rdf\text{-}D} E$ are decidable in LogSpace in the size of $S$, and in the combined size of the graphs if $E$ is ground.*

Table 2 summarizes the complexity of reasoning with the entailment regimes of RDF; "stand. RDFS" stands for "only standard use of the RDFS vocabulary; $S$ and $E$ are such that $E \trianglelefteq S$". The results in the first and fourth line of the table, and the upper bound for ground *rdfs*-entailment were previously known (Gutierrez et al., 2004; de Bruijn et al., 2005; ter Horst, 2005). To the best of our knowledge, the other results are novel.

Table 3 summarizes the complexity of reasoning with datatypes, measured in the size of the entailing graph $S$. "Definite $D$" stands for $D$ entailment restricted to definite datatype maps. The LogSpace results require the datatype map $D$ to be logspace as well, i.e., it must be decidable in LogSpace whether two literals are equal under the interpretation given by $D$. We suspect that many datatype maps of interest are logspace – examples are the XML schema datatypes (Peterson et al., 2009). The upper bounds for *rdfs*- and *rdfs*-$D^*$- entailment are known from the literature (ter Horst, 2005). To the best of our knowledge, the other results in the table are novel.

## 6. Conclusions and Future Work

We have presented embeddings of the different RDF entailment regimes in F-Logic, and we have shown how deductive database and description logic technology can be used for reasoning with RDF.

Known complexity results from the fields of deductive databases and description logics resulted in several novel upper bounds, in particular, ground simple- and *rdf*-entailment are in LogSpace, as are the respective extensions with $D^*$ datatype semantics; non-ground (resp., ground) *erdfs*-entailment of graphs with only standard use of the RDFS vocabulary is in NP (resp., NLogSpace). To the best of our knowledge these are the first known upper bounds for extensional RDFS entailment for a nontrivial subset of RDF graphs. For the case of extensions of simple-, *rdf*-, and *rdfs*-entailment with $D$ datatype support, the upper bounds for non-ground and ground entailment are the same as for $D^*$ entailment when considering *definite* datatypes, which do not require reasoning by case.

In addition, we have established several lower bounds through reductions from known hard problems. In particular, *rdfs*-entailment turns out to be PTime-hard and simple-entailment extended with $D$ datatype support turns out to be coNP-hard, both in the size of the entailing graph. We also found a matching lower bound for the NLogSpace result for ground *erdfs*-entailment of graphs with only standard use of the RDFS vocabulary.

The negative result concerning ground *rdfs*-entailment (i.e., PTime-hardness) might come as a surprise because the language seems far less expressive than other PTime-hard languages (e.g., variable-free Datalog (Dantsin et al., 2001) and $DL\text{-}Lite_{\mathcal{R},\sqcap}$, an extension of $DL\text{-}Lite_{\mathcal{R}}$ (Calvanese et al., 2007)). The PTime-hardness proof suggests that the complexity originates from the possibility to use RDFS vocabulary in arbitrary places in RDF statements, e.g., `rdfs:subPropertyOf` in the object position of a triple. Indeed, ground entailment in the minimal RDFS fragment by Muñoz, Pérez, and Gutierrez (2009) can be decided in $O(nlogn)$.[6] We suspect that the minimal RDFS fragment can be extended with many useful features, such as class and property declarations and the RDFS metadata vocabulary, without compromising the $O(nlogn)$ upper bound. This is a topic for future work.

The negative result concerning $D$ entailment, even when not considering the RDFS vocabulary (i.e., coNP-hardness), suggests that one should restrict oneself to a weaker datatype semantics such as $D^*$ or one should use only definite datatype maps, which precludes the use of finite datatypes such as `bool` or `int` (Peterson et al., 2009). The latter approach was taken in the specification of the tractable fragments (also called *profiles*) of OWL 2 (Motik et al., 2009a), which has a datatype semantics similar to the $D$ semantics.

The investigation reported on in this paper has formed the basis for the specification of combinations of RIF-BLD rules (RIF Working Group, 2010a), which are essentially Horn logic formulas, with RDF graphs. The RIF RDF and OWL specification (RIF Working Group, 2010b) gives a model-theoretic account of the semantics of RIF-RDF combinations

6. This minimal RDFS disallows the use of any RDF(S) vocabulary besides the properties in the RDF(S) ontology vocabulary, and allows the use of these properties only in the predicate position of triples.

and suggests how such combinations can be embedded into RIF-BLD rules, based on the embedding in Section 3.2. A particular challenge for future work is the combination of RDF graphs with extensions of RIF-BLD that allow nonmonotonic negation in the rules, and the interaction of this negation with blank nodes.

Another topic for future investigation is the precise relationship between extensional RDFS and OWL. In particular, the relationship between extensional RDFS with only standard use of the RDFS vocabulary and the OWL 2 QL fragment of OWL 2 (Motik et al., 2009a), which is based on contextual $DL\text{-}Lite_{\mathcal{R}}$. The embedding in the proof of Proposition 7 provides a promising starting point.

## Acknowledgments

Jos de Bruijn was partially supported by the European Commission under the projects Knowledge Web (IST-2004-507482) and ONTORULE (FP7 231875). Stijn Heymans was partially supported by the Austrian Science Fund (FWF) under projects P20305 and P20840 and by the European Commission under the project ONTORULE (FP7 231875).

## Appendix A. RDF(S) Semantics

In this appendix we define the notions of RDF, RDFS, eRDFS, $D^*$, and $D$ interpretations (Hayes, 2004; ter Horst, 2005). Recall the definition of interpretation in Section 3.1.

**RDF Interpretations** The RDF vocabulary consists of the following symbols:

`rdf:type` `rdf:Property` `rdf:XMLLiteral` `rdf:nil` `rdf:List` `rdf:Statement` `rdf:subject`
`rdf:predicate` `rdf:object` `rdf:first` `rdf:rest` `rdf:Seq` `rdf:Bag` `rdf:value` `rdf:Alt`
`rdf:_1` `rdf:_2` ...

The *RDF ontology vocabulary* consists of the symbols `rdf:type` and `rdf:Property`. Note that `rdf:_i`, for any positive integer `i`, is part of the RDF vocabulary. Thus, the RDF vocabulary is infinite. In the remainder, we omit the prefix `rdf:` when using the RDF vocabulary.

A typed literal $(s, \texttt{XMLLiteral})$ is a *well-typed XML literal* if $s$ is in the lexical space of `XMLLiteral`, as defined in (Klyne & Carroll, 2004, Section 5.1); the *XML value* of $s$, denoted $xml(s)$, is in one-to-one correspondence with $s$. If $s$ is not in the lexical space of `XMLLiteral`, then $(s, \texttt{XMLLiteral})$ is an *ill-typed XML literal.*

Given an interpretation $I$, the *class extension* of an object $x \in IR$ is the set of elements connected to $x$ via `type`, i.e., the instances of $x$. It is defined as $ICEXT(x) = \{k \mid \langle k, x\rangle \in IEXT(IS(\texttt{type}))\}$.

An interpretation $I$ of a vocabulary $V = \langle \mathcal{C}, \mathcal{PL}, \mathcal{TL}\rangle$ is an *rdf-interpretation* if $V$ includes the RDF vocabulary and conditions 1–4 in Table 4 hold in $I$.

**RDFS Interpretations** The RDFS vocabulary consists of:

`rdfs:domain` `rdfs:range` `rdfs:Resource` `rdfs:Literal` `rdfs:Datatype` `rdfs:Class`
`rdfs:subClassOf` `rdfs:subPropertyOf` `rdfs:member` `rdfs:Container` `rdfs:label`
`rdfs:ContainerMembershipProperty` `rdfs:comment` `rdfs:seeAlso` `rdfs:isDefinedBy`

| | |
|---|---|
| 1 | $IS(\texttt{type}), IS(\texttt{subject}), IS(\texttt{predicate}), IS(\texttt{object}), IS(\texttt{first}), IS(\texttt{rest}),$ $IS(\texttt{value}), IS(\texttt{_1}), IS(\texttt{_2}), \ldots \in IP$ |
| | $IS(\texttt{nil}) \in ICEXT(IS(\texttt{List}))$ |
| 2 | $IP = ICEXT(IS(\texttt{Property}))$ |
| 3 | if $(s, \texttt{XMLLiteral}) \in \mathcal{TL}$ is a well-typed XML literal, then $IL((s, \texttt{XMLLiteral})) = xml(s)$, $IL((s, \texttt{XMLLiteral})) \in LV$, and $IL((s, \texttt{XMLLiteral})) \in ICEXT(IS(\texttt{XMLLiteral}))$ |
| 4 | if $(s, \texttt{XMLLiteral}) \in \mathcal{TL}$ is an ill-typed XML literal, then $IL((s, \texttt{XMLLiteral})) \notin LV$ and $IL((s, \texttt{XMLLiteral})) \notin ICEXT(IS(\texttt{XMLLiteral}))$ |

Table 4: Conditions on RDF interpretations

The *RDFS ontology vocabulary* consists of the symbols `rdfs:subClassOf`, `rdfs:subPropertyOf`, `rdfs:domain`, `rdfs:range`, `rdfs:Class`, and `rdfs:Datatype`. In the remainder we omit the prefix `rdfs:` when using the RDFS vocabulary.

We say that an *rdf*-interpretation $I$ of a vocabulary $V$ is an *rdfs-interpretation* if $V$ includes the RDFS vocabulary and conditions 5–15 depicted in Table 5 hold in $I$. As a shortcut, we define $IEXT_p(o) = \{s \mid \langle s, IS(o)\rangle \in IEXT(IS(p))\}$.

An RDF (resp, RDFS) axiomatic triple is a triple that is satisfied in every *rdf*-(resp, rdfs-) interpretation. Conditions 1 and 5 correspond to the RDF(S) axiomatic triples in the following way; see also (Hayes, 2004, Sections 3.1 and 4.1):

- $IS(s) \in IP$ corresponds to the axiomatic triple $\langle s, \texttt{type}, \texttt{rdf:Property}\rangle$,
- $IS(s) \in IEXT_p(o)$ corresponds to the axiomatic triple $\langle s, p, o\rangle$, and
- $IS(s) \in ICEXT(IS(c))$ corresponds to the axiomatic triple $\langle s, \texttt{type}, c\rangle$.

**Extensional RDFS Interpretations** The normative RDFS semantics, reviewed above, is also called the *intensional* RDFS semantics. The RDF semantics specification (Hayes, 2004) also defines an *extensional* RDFS semantics (eRDFS).

An *rdfs*-interpretation $I$ is an *erdfs-interpretation* if the conditions depicted in Table 6 hold.

***D** Interpretations** Given a vocabulary $V$ and a simple datatype map $D$, an $s$-interpretation of $V$ is an *s-D*-interpretation* if $V$ includes $dom(D)$ and conditions 16–19 in Table 7 are satisfied for each $u \in dom(D)$.

Given a vocabulary $V$ and datatype map $D$, an *rdf* (resp., *rdfs*, *erdfs*)-interpretation $I$ of $V$ is an *rdf-D** (resp., *rdfs-D**, *erdfs-D**)-interpretation if $I$ is an *s-D**-interpretation.

***D* Interpretations** Given a vocabulary $V$ and a simple datatype map $D$, an *s-D**-interpretation of $V$ is an *s-D-interpretation* if it satisfies conditions 20–22 in Table 8 for each $u \in dom(D)$.

Given a vocabulary $V$ and a datatype map $D$, an *rdf* (resp., *rdfs*, *erdfs*)-interpretation $I$ of $V$ is an *rdf-D* (resp., *rdfs-D*, *erdfs-D*)-interpretation if $I$ is an *s-D*-interpretation.

| | |
|---|---|
| 5 | $IS$(`type`), $IS$(`member`), $IS$(`seeAlso`), $IS$(`isDefinedBy`), $IS$(`comment`), $IS$(`label`), $IS$(`value`), $IS$(`_1`), $IS$(`_2`), $\ldots \in IEXT_{\texttt{domain}}$(`Resource`) |
| | $IS$(`domain`), $IS$(`range`), $IS$(`subPropertyOf`) $\in IEXT_{\texttt{rdfs:domain}}$(`Property`) |
| | $IS$(`subClassOf`) $\in IEXT_{\texttt{rdfs:domain}}$(`Class`) |
| | $IS$(`subject`), $IS$(`predicate`), $IS$(`object`) $\in IEXT_{\texttt{domain}}$(`Statement`) |
| | $IS$(`first`), $IS$(`rest`) $\in IEXT_{\texttt{domain}}$(`List`) |
| | $IS$(`subject`), $IS$(`predicate`), $IS$(`object`), $IS$(`member`), $IS$(`first`), $IS$(`seeAlso`), $IS$(`isDefinedBy`), $IS$(`value`), $IS$(`_1`), $IS$(`_2`), $\ldots \in IEXT_{\texttt{range}}$(`Resource`) |
| | $IS$(`comment`), $IS$(`label`) $\in IEXT_{\texttt{range}}$(`Literal`) |
| | $IS$(`subPropertyOf`) $\in IEXT_{\texttt{range}}$(`Property`) |
| | $IS$(`type`), $IS$(`domain`), $IS$(`range`), $IS$(`subClassOf`) $\in IEXT_{\texttt{range}}$(`Class`) |
| | $IS$(`rest`) $\in IEXT_{\texttt{range}}$(`List`) |
| | $IS$(`Alt`), $IS$(`Bag`), $IS$(`Seq`) $\in IEXT_{\texttt{subClassOf}}$(`Container`) |
| | $IS$(`ContainerMembershipProperty`) $\in IEXT_{\texttt{subClassOf}}$(`Property`) |
| | $IS$(`isDefinedBy`) $\in IEXT_{\texttt{subPropertyOf}}$(`seeAlso`) |
| | $IS$(`XMLLiteral`) $\in ICEXT(IS$(`Datatype`)) |
| | $IS$(`XMLLiteral`) $\in IEXT_{\texttt{subClassOf}}$(`Literal`) |
| | $IS$(`Datatype`) $\in IEXT_{\texttt{subClassOf}}$(`Class`) |
| | $IS$(`_1`), $IS$(`_2`), $\ldots \in ICEXT(IS$(`ContainerMembershipProperty`)) |
| 6 | $IR = ICEXT(IS$(`Resource`))<br>$LV = ICEXT(IS$(`Literal`)) |
| 7 | if $\langle x, y\rangle \in IEXT(IS$(`domain`)) and $\langle u, v\rangle \in IEXT(x)$, then $u \in ICEXT(y)$ |
| 8 | if $\langle x, y\rangle \in IEXT(IS$(`range`)) and $\langle u, v\rangle \in IEXT(x)$, then $v \in ICEXT(y)$ |
| 9 | $IEXT(IS$(`subPropertyOf`)) is transitive and reflexive on $IP$ |
| 10 | if $\langle x, y\rangle \in IEXT(IS$(`subPropertyOf`)), then $IEXT(x) \subseteq IEXT(y)$ |
| 11 | if $x \in ICEXT$(`Class`), then $x \in IEXT_{\texttt{subClassOf}}$(`Resource`) |
| 12 | if $\langle x, y\rangle \in IEXT(IS$(`subClassOf`)), then $ICEXT(x) \subseteq ICEXT(y)$ |
| 13 | $IEXT(IS$(`subClassOf`)) is transitive and reflexive on $ICEXT$(`Class`) |
| 14 | if $x \in ICEXT$(`ContainerMembershipProperty`), then $x \in IEXT_{\texttt{subPropertyOf}}$(`member`) |
| 15 | if $x \in ICEXT$(`Datatype`), then $x \in IEXT_{\texttt{subClassOf}}$(`Literal`) |

Table 5: Conditions on RDFS interpretations

| | |
|---|---|
| 7′ | $\langle x, y\rangle \in IEXT(IS$(`domain`)) if and only if (if $\langle u, v\rangle \in IEXT(x)$, then $u \in ICEXT(y)$) |
| 8′ | $\langle x, y\rangle \in IEXT(IS$(`range`)) if and only if (if $\langle u, v\rangle \in IEXT(x)$, then $v \in ICEXT(y)$) |
| 10′ | $\langle x, y\rangle \in IEXT(IS$(`subPropertyOf`)) if and only if $x, y \in IP$ and $IEXT(x) \subseteq IEXT(y)$ |
| 12′ | $\langle x, y\rangle \in IEXT(IS$(`subClassOf`)) if and only if<br>$x, y \in ICEXT$(`Class`) and $ICEXT(x) \subseteq ICEXT(y)$ |

Table 6: Conditions on eRDFS interpretations

| | |
|---|---|
| 16 | $IS(u) = D(u)$ |
| 17 | $IS(u) \in ICEXT(IS(\texttt{Datatype}))$ |
| 18 | if $(s, u) \in \mathcal{TL}$ and $s \in L^{D(u)}$, then $IL((s, u)) = L2V^{D(u)}(s) \in LV$ and $IL((s, u)) \in ICEXT(D(u))$ |
| 19 | if $(s, u) \in \mathcal{TL}$ and $s \notin L^{D(u)}$, then $IL((s, u)) \notin LV$ and $IL((s, u)) \notin ICEXT(D(u))$ |

Table 7: Conditions on $D^*$ interpretations

| | |
|---|---|
| 20 | $ICEXT(IS(u)) = V^{D(u)} \subseteq LV$ |
| 21 | if $(s, u') \in \mathcal{TL}$, $IS(u') = IS(u)$ and $s \in L^{D(u)}$, then $IL((s, u')) = L2V^{D(u)}(s)$ |
| 22 | if $(s, u') \in \mathcal{TL}$, $IS(u') = IS(u)$ and $s \notin L^{D(u)}$, then $IL((s, u')) \notin LV$ |

Table 8: Conditions on $D$-interpretations

## Appendix B. Embeddings

This appendix contains the axiomatization $\Psi^x$ of the entailment regimes $x \in \{s, rdf, rdf, erdfs\}$ and the axiomatization of the datatype entailment regimes $\Psi^{x\text{-}D^*}$, $\Psi^{x\text{-}D}$, referenced from Sections 3 and 4.

Following the convention in Appendix A we omit the prefixes `rdf:` and `rdfs:` when using the RDF and RDF vocabularies.

### B.1 RDF Entailment Regimes

The axiomatization of the $s$, $rdf$, $rdfs$, and $erdfs$ entailment regimes, denoted $\Psi^x$, for $x \in \{s, rdf, rdfs, erdfs\}$, is defined in Table 9.

### B.2 Datatype Entailment Regimes

The axiomatization of the $D^*$ and $D$ entailment regimes, denoted $\Psi^{x\text{-}D^*}$ and $\Psi^{x\text{-}D}$, respectively, for $x \in \{s, rdf, rdfs, erdfs\}$, is defined in Table 10.

Note that $D$ entailment requires that whenever two URIs are mapped to the same individual in a given interpretation, the URIs can be used interchangeably in typed literals. However, since equality between URIs cannot be stated in RDF(S) – or indeed inferred – we do not need to consider this case in our embeddings.

### B.3 Extensional RDFS

Let $V = \langle \mathcal{C}, \mathcal{PL}, \mathcal{TL} \rangle$ be a vocabulary. The mapping function $tr^{erdfs}$, defined in Section 3.3, deals with the eRDFS semantics of most of the RDF(S) vocabulary through direct embedding. We define here the theory $\Psi^{erdfs\text{-}V}$, which deals with the remainder of the RDF(S) vocabulary.

$$\Psi^{s} = \emptyset$$

$$\begin{aligned}
\Psi^{rdf} = {} & \{tr(\langle s,p,o\rangle) \mid \langle s,p,o\rangle \text{ is an RDF axiomatic triple}\}\cup \\
& \{t[\mathtt{type} \twoheadrightarrow \mathtt{XMLLiteral}] \mid t \in \mathcal{TL} \text{ is a well-typed XML literal}\}\cup \\
& \{illD(t) \mid t \in \mathcal{TL} \text{ is an ill-typed XML literal}\}\cup \\
& \{\forall x(\exists y, z(y[x \twoheadrightarrow z]) \supset x[\mathtt{type} \twoheadrightarrow \mathtt{Property}]), \\
& \forall x(x[\mathtt{type} \twoheadrightarrow \mathtt{XMLLiteral}] \wedge illD(x) \supset \bot)\}
\end{aligned}$$

$$\begin{aligned}
\Psi^{rdfs} = {} & \Psi^{rdf} \cup \{tr(\langle s,p,o\rangle) \mid \langle s,p,o\rangle \text{ is an RDFS axiomatic triple}\}\cup \\
& \{t[\mathtt{type} \twoheadrightarrow \mathtt{Literal}] \mid t \in \mathcal{PL}\}\cup \\
& \{\forall x(x[\mathtt{type} \twoheadrightarrow \mathtt{Resource}]), \\
& \forall u,v,x,y(x[\mathtt{domain} \twoheadrightarrow y] \wedge u[x \twoheadrightarrow v] \supset u[\mathtt{type} \twoheadrightarrow y]), \\
& \forall u,v,x,y(x[\mathtt{range} \twoheadrightarrow y] \wedge u[x \twoheadrightarrow v] \supset v[\mathtt{type} \twoheadrightarrow y]), \\
& \forall x(x[\mathtt{type} \twoheadrightarrow \mathtt{Property}] \supset x[\mathtt{subPropertyOf} \twoheadrightarrow x]), \\
& \forall x,y,z(x[\mathtt{subPropertyOf} \twoheadrightarrow y] \wedge y[\mathtt{subPropertyOf} \twoheadrightarrow z] \supset x[\mathtt{subPropertyOf} \twoheadrightarrow z]), \\
& \forall x,y(x[\mathtt{subPropertyOf} \twoheadrightarrow y] \supset \forall z_1,z_2(z_1[x \twoheadrightarrow z_2] \supset z_1[y \twoheadrightarrow z_2])), \\
& \forall x(x[\mathtt{type} \twoheadrightarrow \mathtt{Class}] \supset x[\mathtt{subClassOf} \twoheadrightarrow \mathtt{Resource}]), \\
& \forall x,y(x[\mathtt{subClassOf} \twoheadrightarrow y] \supset \forall z(z[\mathtt{type} \twoheadrightarrow x] \supset z[\mathtt{type} \twoheadrightarrow y])), \\
& \forall x(x[\mathtt{type} \twoheadrightarrow \mathtt{Class}] \supset x[\mathtt{subClassOf} \twoheadrightarrow x]), \\
& \forall x,y,z(x[\mathtt{subClassOf} \twoheadrightarrow y] \wedge y[\mathtt{subClassOf} \twoheadrightarrow z] \supset x[\mathtt{subClassOf} \twoheadrightarrow z]), \\
& \forall x(x[\mathtt{type} \twoheadrightarrow \mathtt{ContainerMembershipProperty}] \supset x[\mathtt{subPropertyOf} \twoheadrightarrow \mathtt{member}]), \\
& \forall x(x[\mathtt{type} \twoheadrightarrow \mathtt{Datatype}] \supset x[\mathtt{subClassOf} \twoheadrightarrow \mathtt{Literal}]), \\
& \forall x(x[\mathtt{type} \twoheadrightarrow \mathtt{Literal}] \wedge illD(x) \supset \bot)\}
\end{aligned}$$

$$\begin{aligned}
\Psi^{erdfs} = {} & \Psi^{rdfs} \cup \{\forall x,y(\forall u,v(u[x \twoheadrightarrow v] \supset u[\mathtt{type} \twoheadrightarrow y]) \supset x[\mathtt{domain} \twoheadrightarrow y]), \\
& \forall x,y(\forall u,v(u[x \twoheadrightarrow v] \supset v[\mathtt{type} \twoheadrightarrow y]) \supset x[\mathtt{range} \twoheadrightarrow y]), \\
& \forall x,y(x[\mathtt{type} \twoheadrightarrow \mathtt{Property}] \wedge y[\mathtt{type} \twoheadrightarrow \mathtt{Property}] \wedge \forall u,v(u[x \twoheadrightarrow v] \supset \\
& \quad u[y \twoheadrightarrow v]) \supset x[\mathtt{subPropertyOf} \twoheadrightarrow y]), \\
& \forall x,y(x[\mathtt{type} \twoheadrightarrow \mathtt{Class}] \wedge y[\mathtt{type} \twoheadrightarrow \mathtt{Class}] \wedge \forall u(u[\mathtt{type} \twoheadrightarrow x] \supset u[\mathtt{type} \twoheadrightarrow y]) \supset \\
& \quad x[\mathtt{subClassOf} \twoheadrightarrow y])\}
\end{aligned}$$

Table 9: Axiomatization of the RDF entailment regimes.

$$\begin{aligned}
\Psi^{erdfs\text{-}V} = {} & \{tr^{erdfs}(\langle s,p,o\rangle) \mid \langle s,p,o\rangle \text{ is an RDF(S) axiomatic triple with} \\
& \quad \text{only standard use of the RDF(S) vocabulary}\}\cup \\
& \{t\,{:}\,\mathtt{XMLLiteral} \mid t \in \mathcal{TL} \text{ is a well-typed XML literal}\}\cup \\
& \{t\,{:}\,illxml \mid t \in \mathcal{TL} \text{ is an ill-typed XML literal}\}\cup \\
& \{t\,{:}\,\mathtt{Literal} \mid t \in \mathcal{PL}\}\cup \\
& \{\forall x(x\,{:}\,\mathtt{Literal} \wedge x\,{:}\,illxml \supset \bot)\}
\end{aligned}$$

## Appendix C. Proofs

This appendix contains the proofs of the propositions and theorems in Sections 3 and 4.

### C.1 Proof of Proposition 2

Consider a generalized RDF graph $S$, an interpretation $I = \langle IR, IP, IS, IEXT\rangle$ for which holds that $IP = IR$ includes every term in $S$, $IS(c) = c$ for any URI $c$, $IL(l) = l$ for any typed literal $l$, and for every triple $\langle s,p,o\rangle \in S$, $(s,o) \in IEXT(p)$, and a blank node

$$\Psi^{V\text{-}D^*\text{-}=} = \{l = (s,u) \mid l \in \mathcal{PL},\ (s,u) \in \mathcal{TL} \text{ is a well-typed literal, and } l = L2V^{D(u)}(s)\} \cup \{(s,u) = (s',u') \mid (s,u),(s',u') \in \mathcal{TL} \text{ are distinct well-typed literals and } L2V^{D(u)}(s) = L2V^{D(u')}(s')\}$$

$$\Psi^{x\text{-}D^*} = \Psi^x \cup \Psi^{V\text{-}D^*\text{-}=} \cup \{(s,u)[\mathtt{type} \twoheadrightarrow u] \mid (s,u) \in \mathcal{TL} \text{ is a well-typed literal}\} \cup \{illD(t) \mid t \in \mathcal{TL} \text{ is an ill-typed literal}\} \cup \{u[\mathtt{type} \twoheadrightarrow \mathtt{Datatype}] \mid u \in dom(D)\} \cup \{\forall x(illD(x) \wedge x[\mathtt{type} \twoheadrightarrow u] \supset \bot) \mid u \in dom(D)\}$$

$$\Psi^{x\text{-}D} = \Psi^{x\text{-}D^*} \cup \{(s,u')[\mathtt{type} \twoheadrightarrow u] \mid (s,u') \in \mathcal{TL} \text{ is a well-typed literal, } u \in dom(D), \text{ and } L2V^{D(u')}(s) \in V^{D(u)}\} \cup \{s[\mathtt{type} \twoheadrightarrow u] \mid s \in \mathcal{PL}, u \in dom(D), \text{ and } s \in V^{D(u)}\} \cup \{\forall x(x[\mathtt{type} \twoheadrightarrow u] \supset dt(x,u)) \mid u \in dom(D)\} \cup \{\exists x(dt(x,u_1) \wedge dt(x,u_2)) \supset \bot \mid u_1,u_2 \in dom(D).V^{D(u_1)} \cap V^{D(u_2)} = \emptyset\} \cup \{dt(l,u) \supset \bot \mid l \in \mathcal{PL}, u \in dom(D), l \notin V^{D(u)}\} \cup \{dt((s,u),u') \supset \bot \mid (s,u) \in \mathcal{TL}, u' \in dom(D), L2V^{D(u)}(s) \notin V^{D(u')}\}$$

Table 10: Axiomatization of the datatype entailment regimes, $x \in \{s, rdf, rdfs, erdfs\}$.

assignment $A : bl(S) \to IR$ that maps every blank node to itself. Clearly, $(I, A) \models S$, $I \models S$, and $I$ is an $s$-interpretation. Therefore, $S$ is $s$-satisfiable.

It is easy to see that the following generalized RDF graph is *rdf*-, and hence *rdfs*- and *erdfs*-unsatisfiable, by the negation in condition 4 in Table 4: $S = \{\langle(\text{“<notXML”}, \mathtt{XMLLiteral}), \mathtt{type}, \mathtt{XMLLiteral}\rangle\}$: $(\text{“<notXML”}, \mathtt{XMLLiteral})$ is an ill-typed XML literal, so by condition 4 in Table 4, $IL((\text{“<notXML”}, \mathtt{XMLLiteral})) \notin ICEXT(IS(\mathtt{XMLLiteral}))$. However, if the graph were satisfied in some *rdf*-interpretation, it must be the case that $IL((\text{“<notXML”}, \mathtt{XMLLiteral})) \in ICEXT(IS(\mathtt{XMLLiteral}))$, a contradiction.

Hayes (2004) observed that one can create a similar situation with a normal RDF graph and a `range` constraint; this graph is *rdfs*- and hence *erdfs*-unsatisfiable. □

## C.2 Proof of Theorem 1

We first show that $S \not\models_x E$ iff $tr(S) \cup \Psi^x \not\models_{\mathsf{f}} tr(E)$. From this follows immediately that $S \models_x E$ iff $tr(S) \cup \Psi^x \models_{\mathsf{f}} tr(E)$.

($\Rightarrow$) Let $V = \langle \mathcal{C}, \mathcal{PL}, \mathcal{TL} \rangle$ be the vocabulary of $S$ and $E$ and let $\mathcal{L}$ be an F-language that conforms with $V$. Assume that $S \not\models_x E$. This means that there is an $x$-interpretation $I = \langle IR, IP, LV, IS, IL, IEXT \rangle$ such that $I \models S$ and $I \not\models E$. We construct a corresponding F-structure $\mathbf{I} = \langle U, \in_U, \mathbf{I}_C, \mathbf{I}_{\twoheadrightarrow}, \mathbf{I}_P \rangle$ in the following way:

(i) $U = IR \cup IP$,

(ii) $\mathbf{I}_F(t) = IS(t)$ for every URI reference $t \in \mathcal{C}$, $\mathbf{I}_F(t) = t$ for every plain literal $t \in \mathcal{PL}$, and $\mathbf{I}_F(t) = IL(t)$ for every typed literal $t \in \mathcal{TL}$,

(iii) $\mathbf{I}_{\twoheadrightarrow}(k) = IEXT(k)$ for every $k \in IP$ and $\mathbf{I}_{\twoheadrightarrow}(k) = \emptyset$ for every $k \notin IP$,

(iv) $\mathbf{I}_P(illD) = \{u \mid t \in \mathcal{TL}$ is an ill-typed XML literal and $IL(t) = u\}$.

It is straightforward to verify that $\mathbf{I} \models_{\mathsf{f}} tr(S) \cup \Psi^x$ and $\mathbf{I} \not\models_{\mathsf{f}} tr(E)$. Hence, $tr(S) \cup \Psi^x \not\models_{\mathsf{f}} tr(E)$.

($\Leftarrow$) Assume that $tr(S) \cup \Psi^x \not\models_{\mathsf{f}} tr(E)$. This means that (by (Fitting, 1996, Theorem 5.9.4) and Proposition 6) there is a Herbrand F-structure $\mathbf{I} = \langle U, \in_U, \mathbf{I}_C, \mathbf{I}_{\twoheadrightarrow}, \mathbf{I}_P\rangle$ such that $\mathbf{I} \models_{\mathsf{f}} tr(S) \cup \Psi^x$ and $\mathbf{I} \not\models_{\mathsf{f}} tr(E)$. Since $\mathbf{I}$ is a Herbrand F-structure, $U$ includes all constant symbols, and $\mathbf{I}_C$ maps every constant symbol to itself. We construct a corresponding interpretation $I = \langle IR, IP, LV, IS, IL, IEXT\rangle$ as follows:

(i) $IP = \{p \mid \langle p, \mathbf{I}_F(\texttt{Property})\rangle \in \mathbf{I}_{\twoheadrightarrow}(\mathbf{I}_F(\texttt{type}))\} \cup \{p \mid \exists s, o.\langle s, o\rangle \in \mathbf{I}_{\twoheadrightarrow}(p)\}$,

(ii) $LV = \mathcal{PL} \cup \{xml(s) \mid ((s, \texttt{XMLLiteral}) \in \mathcal{TL} \wedge (s, \texttt{XMLLiteral})$ is a well-typed XML literal$)\} \cup \{l \mid \langle l, \mathbf{I}_F(\texttt{Literal})\rangle \in \mathbf{I}_{\twoheadrightarrow}(\mathbf{I}_F(\texttt{type}))\}$,

(iii) $IR = U \cup LV$,

(iv) $IS(t) = \mathbf{I}_F(t)$ for every URI $t \in \mathcal{C}$, $IL((s, u)) = xml(s)$ if $(s, u) \in \mathcal{TL}$ is a well-typed XML literal; $IL((s, u)) = \mathbf{I}_F((s, u))$ for $(s, u) \in \mathcal{TL}$ if $(s, u) \in \mathcal{TL}$ is not a well-typed XML literal, and

(v) for any $p \in U$ and any $\langle s, o\rangle \in \mathbf{I}_{\twoheadrightarrow}(p)$: $\langle s', o'\rangle \in IEXT(p)$, where $s'$ (resp., $o'$) is: if there is some $(t, \texttt{XMLLiteral}) \in \mathcal{TL}$ such that $(t, \texttt{XMLLiteral})$ is a well-typed XML literal and $\mathbf{I}_F((t, \texttt{XMLLiteral})) = s$ (resp., $\cdots = o$), then $s' = xml(t)$ (resp., $o' = xml(t)$); otherwise $s' = s$ (resp., $o' = o$).

One can verify that $I$ is an $x$-interpretation, $I \models S$, and $I \not\models E$. Hence, $S \not\models E$.

The second part of the theorem can be shown analogously. □

### C.3 Proof of Proposition 4

By Corollary 1 we have that $S \models_x E$ iff $sk(tr(S)) \cup \Psi^x \models_{\mathsf{f}} tr(E)$. Therefore, we need to show $sk(tr(S)) \cup \Psi^x_{-S\cup E} \models_{\mathsf{f}} tr(E)$ iff $sk(tr(S)) \cup \Psi^x \models_{\mathsf{f}} tr(E)$.
($\Rightarrow$) Trivial, since $sk(tr(S)) \cup \Psi^x_{-S\cup E} \subseteq sk(tr(S)) \cup \Psi^x$.
($\Leftarrow$) Consider the case $x = erdfs$. Let $\mathbf{I}$ be a minimal Herbrand model of $sk(tr(S)) \cup \Psi^x$ and let $\mathbf{I}'$ be obtained from $\mathbf{I}$ by removing all triples involving container membership properties $_n$ that appear in $\Psi^x \backslash \Psi^x_{-S\cup E}$. We can verify, e.g., by doing a case analysis on the shape of the triples in $S$, that $\mathbf{I}'$ is a minimal model of $sk(tr(S)) \cup \Psi^x_{-S\cup E}$. Similarly, one can verify that if $\mathbf{I} \models_{\mathsf{f}} tr(E)$, then $\mathbf{I}' \models_{\mathsf{f}} tr(E)$.

Analogous for $x \in \{s, rdf, rdfs\}$. □

### C.4 Proof of Theorem 2

We prove both directions by contraposition.

($\Rightarrow$) Assume $tr^{erdfs}(S) \cup \Psi^{erdfs\text{-}V} \not\models_{\mathsf{f}} tr^{erdfs}(E)$. This means that there is a Herbrand F-structure $\mathbf{I} = \langle U, \in_U, \mathbf{I}_C, \mathbf{I}_{\twoheadrightarrow}, \mathbf{I}_P\rangle$ such that $\mathbf{I} \models_{\mathsf{f}} tr^{erdfs}(S) \cup \Psi^{erdfs\text{-}V}$ and $\mathbf{I} \not\models_{\mathsf{f}} tr^{erdfs}(E)$.

We define $xml'(x) = xml(s)$ if $x$ is a well-typed XML literal $(s, \texttt{XMLLiteral})$; otherwise $xml'(x) = x$. We construct a corresponding interpretation $I = \langle IR, IP, LV, IS, IL, IEXT\rangle$ as follows:

(i) $LV = \{xml'(l) \mid l \in_U \mathbf{I}_F(\texttt{Literal})\}$,

(ii) $IP = IR = U \cup LV \cup \{\texttt{type}, \texttt{subClassOf}, \texttt{domain}, \texttt{range}, \texttt{subPropertyOf}\}$,

(iii) $IS(t) = t$ for every URI reference $t$ in $S$, $E$, or the RDF(S) vocabulary,

(iv) $IL(t) = xml'(t)$ for any $t \in \mathcal{TL}$,

(v) for any $p \in U$ and any $\langle s, o\rangle \in \mathbf{I}_{\twoheadrightarrow}(p)$, $\langle xml'(s), xml'(o)\rangle \in IEXT(p)$,

(vi) $IEXT(\texttt{type})$ is the smallest relation such that

  (a) $IEXT(\texttt{type}) \supseteq \{\langle xml'(s), xml'(o)\rangle \mid s \in_U o\}$;
  (b) $ICEXT(\texttt{Resource}) = ICEXT(\texttt{Class}) = IR$; and
  (c) $ICEXT(\texttt{Property}) = IP$,

(vii) $IEXT(\texttt{domain})$ is the set of all tuples $\langle x, y\rangle$, $x, y \in IR$, such that (if $\langle u, v\rangle \in IEXT(x)$, then $u \in ICEXT(y)$); analogous for $IEXT(\texttt{subClassOf})$, $IEXT(\texttt{subPropertyOf})$, and $IEXT(\texttt{range})$ (see Table 6 for the precise conditions).

One can verify that $I \models S$, and $I \not\models E$, since $S$ and $E$ have only standard use of the RDFS vocabulary, $E$ does not include occurrences of `Resource`, `Class`, or `Property`, and the class and property vocabularies of $E$ are subsets of the respective vocabularies of $S$.

$I$ clearly satisfies conditions 1–4 in Table 4, conditions 6–15 in Table 5, and conditions 7′–12′ in Table 6. To verify that $I$ satisfies condition 5 in Table 4 one only needs to keep in mind that $ICEXT(\texttt{Resource}) = ICEXT(\texttt{Class}) = IR$ and $ICEXT(\texttt{Property}) = IP$. So, $I$ is an *erdfs*-interpretation and thus $S \not\models_{erdfs} E$.

($\Leftarrow$) Assume $S \not\models_{erdfs} E$. This means there is some *erdfs*-interpretation $I'$ such that, for any URI $t$, $IS(t) = t$ (making $I'$ similar to a Herbrand interpretation) and such that $I' \models S$ and $I' \not\models E$. Let $I = \langle IR, IP, LV, IS, IL, IEXT\rangle$ be an *erdfs*-interpretation obtained from $I'$ such that $IP = IR$ and $ICEXT(IS(\texttt{Class})) = IR$, and such that $IEXT$ is minimally extended to satisfy the semantic conditions in Tables 4, 5, and 6. Clearly, there must be such an *erdfs*-interpretation and $I \models S$. We also have, by the restrictions on the class and property vocabularies as well as the non-occurrence in $E$ of `Resource`, `Class`, and `Property`, that $I \not\models E$.

We construct a corresponding F-Logic interpretation $\mathbf{I} = \langle U, \in_U, \mathbf{I}_C, \mathbf{I}_{\twoheadrightarrow}, \mathbf{I}_P\rangle$ as follows: (i) $U = IR$, (ii) $\mathbf{I}_F(t) = t$ for every URI or plain literal $t$, $\mathbf{I}_F(t) = t$ for every $t \in \mathcal{TL}$, (iii) $\mathbf{I}_{\twoheadrightarrow}(k) = IEXT(k)$ for every $k \in U$, and (iv) $\in_U = IEXT(IS(\texttt{type}))$.

It can be straightforwardly verified that $\mathbf{I} \models_{\mathsf{f}} tr^{erdfs}(S) \cup \Psi^{erdfs\text{-}V}$ and $\mathbf{I} \not\models_{\mathsf{f}} tr^{erdfs}(E)$. Therefore, it must be the case that $tr^{erdfs}(S) \cup \Psi^{erdfs\text{-}V} \not\models_{\mathsf{f}} tr^{erdfs}(E)$. $\square$

### C.5 Proof of Proposition 7

$\Phi$ is obtained from $FO(tr^{erdfs}(S) \cup \Psi^{erdfs\text{-}V})$ in the following way:

(i) Class membership and property value statements of the forms $A(a), P(a_1, a_2)$ are included as such,

(ii) Subclass and subproperty statements are included as such,

(iii) Domain constraints of the form $\forall x, y(P(x,y) \supset A(x))$ are rewritten to role-typing statements of the form $\forall x(\exists y(P(x,y)) \supset A(x))$,

(iv) Range constraints of the form $\forall x, y(P(x,y) \supset A(y))$ are rewritten to role-typing statements of the form $\forall x(\exists y(P(y,x)) \supset A(x))$, and

(v) Constraints of the form $\forall x(A(x) \wedge B(x) \supset \bot)$ are rewritten to $\forall x(A(x) \supset \neg B(x))$.

$\Phi$ and $FO(tr^{erdfs}(S))$ are obviously equivalent, and it is easy to verify that $\Phi$ is the FOL equivalent of a contextual $DL\text{-}Lite_{\mathcal{R}}$ knowledge base. □

### C.6 Proof of Theorem 4

We first establish the second part of the theorem, i.e., $S$ is $x$-$D^*$-satisfiable iff $tr(S) \cup \Psi^{x\text{-}D^*}$ has a model.

($\Rightarrow$) Let $V = \langle \mathcal{C}, \mathcal{PL}, \mathcal{TL} \rangle$ be the vocabulary of $S$ and let $\mathcal{L}$ be an F-language that conforms with $V$. Assume that $S$ is $x$-$D^*$-satisfiable. This means that there is an $x$-$D^*$-interpretation $I = \langle IR, IP, LV, IS, IL, IEXT \rangle$ such that $I \models S$. We construct a corresponding F-structure $\mathbf{I} = \langle U, \in_U, \mathbf{I}_C, \mathbf{I}_{\twoheadrightarrow}, \mathbf{I}_P \rangle$ in the following way (analogous to the construction in the '$\Rightarrow$' direction in the proof of Theorem 1):

(i) $U = IR \cup IP$,

(ii) $\mathbf{I}_F(t) = IS(t)$ for every URI reference $t \in \mathcal{C}$, $\mathbf{I}_F(t) = t$ for every plain literal $t \in \mathcal{PL}$, $\mathbf{I}_F(t) = IL(t)$ for every typed literal $t \in \mathcal{TL}$,

(iii) $\mathbf{I}_{\twoheadrightarrow}(k) = IEXT(k)$ for every $k \in IP$,

(iv) $\mathbf{I}_P(illD) = \{u \mid t \in \mathcal{TL}$ is an ill-typed literal and $IL(t) = u\}$.

Clearly, $\mathbf{I} \models_{\mathsf{f}} tr(S)$.

If a literal $t \in \mathcal{TL}$ is an ill-typed XML literal, then clearly it is an ill-typed literal. Then, there is no ill-typed literal $t$ such that $IS(t) \in LV$ (by condition 19 in Table 7), and hence there is no ill-typed literal $t$ such that $IS(t) \in ICEXT(\texttt{Literal})$ or $IS(t) \in ICEXT(\texttt{XMLLiteral})$, by condition 4 in Table 4 (if $x = rdf$, $x = rdfs$, or $x = erdfs$) and condition 6 in Table 5 (if $x = rdfs$ or $x = erdfs$). Satisfaction of $\Psi^x$ is then established straightforwardly.

Consider a well-typed literal $(s, u)$ and a plain literal $l$. In case $l = L2V^{D(u)}(s)$, $IL((s,u)) = l$, by condition 18 in Table 7, and thus $\mathbf{I}_F((s,u)) = \mathbf{I}_F(l) = l$ and $\mathbf{I} \models_{\mathsf{f}} l = (s,u)$, by (ii). Analogous for the case of two distinct well-typed literals. Therefore, $\mathbf{I} \models_{\mathsf{f}} \Psi^{V\text{-}D^*\text{-}=}$.

Consider the definition of $\Psi^{x\text{-}D^*}$ in Table 10. We have established that $\mathbf{I} \models_{\mathsf{f}} \Psi^x \cup \Psi^{V\text{-}D^*\text{-}=}$. Satisfaction of the second, first, third, and fourth sets of formulas in the table follows immediately from, respectively, (iv), and conditions 18, 17, and 19 in Table 7. Therefore, $\mathbf{I} \models_{\mathsf{f}} \Psi^{x\text{-}D^*}$.

This establishes $\mathbf{I} \models_{\mathsf{f}} tr(S) \cup \Psi^{x\text{-}D^*}$.

($\Leftarrow$) Assume that $tr(S) \cup \Psi^{x\text{-}D^*}$ has a model. Let $\Phi = tr(S) \cup \Psi^{x\text{-}D^*}$.

Let $\Phi^{\approx}$ be obtained from $\Phi$ by replacing every occurrence of $=$ with $\approx$ and adding the usual congruence axioms (cf. Fitting, 1996, Chapter 9). It is known that this axiomatization of equality preserves satisfiability and entailment in first-order logic (Fitting, 1996, Theorem 9.3.9). This is also the case for F-Logic, by Proposition 1.

We extend the signature of $\Phi^{\approx}$ with a set of URI references $\mathcal{C}'$, disjoint from $\mathcal{C}$, with cardinality $|bl(S)|$; i.e., the signature is $\Sigma' = \langle \mathcal{C} \cup \mathcal{PL} \cup \mathcal{TL} \cup \mathcal{C}', \mathcal{P} \cup \{\approx\}\rangle$. Since $\Phi^{\approx}$ has a model (as $\Phi$ has), there exists, by classical results, a Herbrand F-structure $\mathbf{I}$ such that $\mathbf{I} \models_{\mathsf{f}} \Phi^{\approx}$. We have that $U = \mathcal{C} \cup \mathcal{C}' \cup \mathcal{PL} \cup \mathcal{TL}$.

For any $u \in U$, define $\sigma$ as follows:

- if $u \in \mathcal{C}$ such that $u \in dom(D)$, $\sigma(u) = D(u)$,
- if $(s, u) \in \mathcal{TL}$ is a well-typed literal and $u \in dom(D)$, $\sigma((s, u)) = L2V^{D(u)}(s)$,
- otherwise, $\sigma(u) = u$.

We construct a corresponding interpretation $I = \langle IR, IP, LV, IS, IL, IEXT\rangle$:

(i) $IP = \{\sigma(p) \mid \langle p, \mathbf{I}_F(\texttt{Property})\rangle \in \mathbf{I}_{\twoheadrightarrow}(\mathbf{I}_F(\texttt{type}))\} \cup \{\sigma(p) \mid \exists s, o.\langle s, o\rangle \in \mathbf{I}_{\twoheadrightarrow}(p)\}$,

(ii) $LV = \mathcal{PL} \cup \{L2V^{D(u)}(s) \mid (s, u) \in \mathcal{TL}, u \in dom(D), (s, u) \text{ is a well-typed literal}\} \cup \{\sigma(l) \mid \langle l, \mathbf{I}_F(\texttt{Literal})\rangle \in \mathbf{I}_{\twoheadrightarrow}(\mathbf{I}_F(\texttt{type})) \;\&\; (x = \mathit{rdfs} \text{ or } x = \mathit{erdfs})\}$,

(iii) $IR = U \cup LV$,

(iv) $IS(u) = \sigma(u)$ for every URI reference $u \in \mathcal{C}$; $IL((s, u)) = \sigma((s, u))$ for every $(s, u) \in \mathcal{TL}$, and

(v) for any $p \in IP$, $IEXT$ is the smallest set such that $\langle s, o\rangle \in \mathbf{I}_{\twoheadrightarrow}(p)$ implies $\langle \sigma(s), \sigma(o)\rangle \in IEXT(\sigma(p))$.

It is easy to see that $I \models S$. Remains to verify that $I$ is an $x$-$D^*$-interpretation. Verifying that $I$ is an $x$-interpretation is straightforward. It remains to verify the satisfaction of the conditions in Table 7.

Satisfaction of condition 16 follows directly from the definition of $I$ and $\sigma$. For condition 17, we have that $u \in dom(D)$ and thus $u[\texttt{type} \twoheadrightarrow \texttt{Datatype}] \in \Psi^{x\text{-}D^*}$. As $\mathbf{I}$ satisfies $\Psi^{x\text{-}D^*}$ we have that $\langle \mathbf{I}_F(u), \mathbf{I}_F(\texttt{Datatype})\rangle \in \mathbf{I}_{\twoheadrightarrow}(\mathbf{I}_F(\texttt{type}))$, and so $\langle \sigma(\mathbf{I}_F(u)), \sigma(\mathbf{I}_F(\texttt{Datatype}))\rangle \in IEXT(\sigma(\mathbf{I}_F(\texttt{type})))$. By construction of $IS$, this yields $IS(u) \in ICEXT(IS(\texttt{Datatype}))$.

Consider some $(s, u) \in \mathcal{TL}$ such that $u \in dom(D)$ and $s \in L^{D(u)}$. Then, $(s, u)$ is well-typed, and so $IL((s, u)) = L2V^{D(u)}(s) \in LV$. By definition of $\Psi^{x\text{-}D^*}$, we have $(s, u)[\texttt{type} \twoheadrightarrow u] \in \Psi^{x\text{-}D^*}$; it follows that $L2V^{D(u)}(s) \in ICEXT(D(u))$. This establishes satisfaction of condition 18.

Condition 19 is satisfied by the fact that $LV$ does not contain ill-typed literals. Indeed, $\mathcal{PL} \cup \{L2V^{D(u)}(s) \mid (s,u) \in \mathcal{TL}, u \in dom(D), (s,u) \text{ is a well-typed literal}\}$ does not contain ill-typed literals and if $x$ is *rdfs* or *erdfs*, there is no ill-typed literal $t$ such that $\mathbf{I} \models_{\mathsf{f}} t[\mathtt{type} \twoheadrightarrow \mathtt{Literal}]$, by the last axiom in the definition of $\Psi^{rdfs}$ in Table 9.

It is easy to verify, for both directions, that we have $I \not\models E$ iff $\mathbf{I} \not\models_{\mathsf{f}} tr(E)$. The first part of the theorem follows. □

### C.7 Proof of Theorem 5

We first show correspondence of satisfiability.

($\Rightarrow$) Let $V = \langle \mathcal{C}, \mathcal{PL}, \mathcal{TL} \rangle$ be the vocabulary of $S$ and $E$, let $rdf$ be an $x$-$D$-interpretation that satisfies $S$ and let $\mathcal{L}$ be an F-language that conforms with $V$. We construct an F-structure $\mathbf{I} = \langle U, \in_U, \mathbf{I}_C, \mathbf{I}_{\twoheadrightarrow}, \mathbf{I}_P \rangle$ that corresponds to $I$, using steps (i)–(iv) as in the ($\Rightarrow$) direction of the proof of Theorem 4, with the additional step

(v) $\mathbf{I}_P(dt) = \{\langle x, u \rangle \mid x \in ICEXT(u) \text{ and } u \in ran(D)\}$.

From the argument in the ($\Rightarrow$) direction in the proof of Theorem 4 follows that $\mathbf{I} \models_{\mathsf{f}} tr(S) \cup \Psi^{x\text{-}D^*}$. Consider $\Psi^{x\text{-}D} \setminus \Psi^{x\text{-}D^*}$, as defined in Table 10. Satisfaction of the first set follows immediately from conditions 20 and 21 in Table 8. Satisfaction of the second set follows immediately from condition 20 in Table 8. Satisfaction of the third set follows immediately from (v).

Consider any two $u_1, u_2 \in dom(D)$ such that $V^{D(u_1)} \cap V^{D(u_2)} = \emptyset$. By condition 20 in Table 8, $ICEXT(IS(u_1)) \cap ICEXT(IS(u_2)) = \emptyset$. From (v) then follows that there is no $k \in U$ such that $\langle k, \mathbf{I}_F(u_1) \rangle \in \mathbf{I}_P(dt)$ and $\langle k, \mathbf{I}_F(u_2) \rangle \in \mathbf{I}_P(dt)$. Consequently, $\mathbf{I} \not\models_{\mathsf{f}} \exists x(dt(x,u_1) \land dt(x,u_2))$ and thus the fourth set of sentences is satisfied.

Consider some $(s,u) \in \mathcal{TL}$ and some $u' \in dom(D)$ such that $IL((s,u)) = L2V^{D(u)}(s) \notin V^{D(u')}$. By condition 20 in Table 8, $ICEXT(IS(u')) = V^{D(u')}$, and thus $IL((s,u)) \notin ICEXT(IS(u'))$. From (v) follows $\langle \mathbf{I}_F((s,u)), \mathbf{I}_F(u') \rangle \notin \mathbf{I}_P(dt)$ and thus $\mathbf{I} \not\models_{\mathsf{f}} dt((s,u),u')$, establishing satisfaction of the sixth set. The argument for the fifth set is obtained by replacing $(s,u) \in \mathcal{TL}$ with $l \in \mathcal{PL}$.

We thus obtain $\mathbf{I} \models_{\mathsf{f}} \Psi^{x\text{-}D}$. Therefore, $tr(S) \cup \Psi^{x\text{-}D}$ has a model.

($\Leftarrow$) Assume $\Phi = tr(S) \cup \Psi^{x\text{-}D}$ has a model.

Let $\Phi^{\approx}$ be obtained from $\Phi$ as in the proof of Theorem 4 and let $\mathbf{I} = \langle U, \in_U, \mathbf{I}_C, \mathbf{I}_{\twoheadrightarrow}, \mathbf{I}_P \rangle$ be a Herbrand F-structure that is a model of $\Phi^{\approx}$. We construct a corresponding interpretation $I = \langle IR, IP, LV, IS, IL, IEXT \rangle$ in the following way. W.l.o.g. we assume that no value space $V^d$, for $d \in ran(D)$, contains any typed literal $t \in \mathcal{TL}$.

We observe that (*) for any two $d_1, d_2 \in dom(D)$ must hold that either $V^{D(d_1)}$ and $V^{D(d_2)}$ are disjoint or their overlap is infinite, since $D$ is definite. In addition, if $V^{D(d_1)}$ and $V^{D(d_2)}$ are disjoint, then, by satisfaction of the fourth set in the definition of $\Psi^{x\text{-}D}$, (**) $\mathbf{I} \not\models_{\mathsf{f}} t[\mathtt{type} \twoheadrightarrow d_1] \land t[\mathtt{type} \twoheadrightarrow d_2]$ for any $t \in \mathcal{C} \cup \mathcal{C}'$.

For a given URI $u \in \mathcal{C} \cup \mathcal{C}'$ we define the mapping $\sigma$ as follows:

- if $u \in dom(D)$, then $\sigma(u) = D(u)$,
- if $\langle u, u' \rangle \in \mathbf{I}_{\twoheadrightarrow}(\mathtt{type})$, for some $u' \in dom(D)$, then $\sigma(u) = v$, where $v$ is such that

- $v \in V^{D(u_1)} \cap \cdots \cap V^{D(u_n)}$, where $u_1, \ldots, u_n \in \mathit{dom}(D)$ are all the datatype identifiers such that $\langle u, u_1 \rangle, \ldots, \langle u, u_n \rangle \in \mathbf{I}_{\twoheadrightarrow}(\mathtt{type})$;
- there is no $u' \in \mathcal{C}$ such that $v = \sigma(u')$; and
- there is no $(s, u') \in \mathcal{TL}$ such that $u' \in \mathit{dom}(D)$ and $v = L2V^{D(u')}(s)$;

such a $v$ must exist, because $u$ cannot be a member of two disjoint datatypes, by (**), and $V^{D(u_1)} \cap \cdots \cap V^{D(u_n)}$ is an infinite set, by Definition 4,

- otherwise, $\sigma(u) = u$.

For a given literal $l \in \mathcal{PL} \cup \mathcal{TL}$ we define $\sigma$ as:

- if $l = (s, u) \in \mathcal{TL}$ is a well-typed literal, then $\sigma(s, u) = L2V^{D(u)}(s)$,
- otherwise $\sigma(l) = \mathbf{I}_F(l)$.

One can verify that $\sigma$ is such that for any two distinct $t_1, t_2 \in \mathcal{C} \cup \mathcal{PL} \cup \mathcal{TL}$, either $\sigma(t_1) = \sigma(t_2)$ and $\langle t_1, t_2 \rangle \in \mathbf{I}_P(\approx)$ (by definition of $\Psi^{V\text{-}D^*\text{-}=}$) or $\sigma(t_1) \neq \sigma(t_2)$.

We construct an RDF interpretation $I = \langle \mathit{IR}, \mathit{IP}, \mathit{LV}, \mathit{IS}, \mathit{IL}, \mathit{IEXT} \rangle$, similar to the construction in the ($\Leftarrow$) direction of the proof of Theorem 4. Note that the respective constructions differ only in steps (ii) and (v).

(i) $\mathit{IP} = \{\sigma(p) \mid \langle p, \mathbf{I}_F(\mathtt{Property}) \rangle \in \mathbf{I}_{\twoheadrightarrow}(\mathbf{I}_F(\mathtt{type}))\} \cup \{\sigma(p) \mid \exists s, o. \langle s, o \rangle \in \mathbf{I}_{\twoheadrightarrow}(p)\}$,

(ii) $\mathit{LV} = \mathcal{PL} \cup \bigcup\{V^d \mid d \in \mathit{ran}(D)\} \cup \{\sigma(l) \mid \langle l, \mathbf{I}_F(\mathtt{Literal}) \rangle \in \mathbf{I}_{\twoheadrightarrow}(\mathbf{I}_F(\mathtt{type}))$ & $(x = \mathit{rdfs}$ or $x = \mathit{erdfs})\}$,

(iii) $\mathit{IR} = U \cup \mathit{LV}$,

(iv) $\mathit{IS}(u) = \sigma(u)$ for every $u \in \mathcal{C}$; $\mathit{IL}((s, u)) = \sigma((s, u))$ for every $(s, u) \in \mathcal{TL}$, and

(v) $\mathit{IEXT}$ is the smallest set such that
- $\mathit{ICEXT}(\mathit{IS}(u)) = V^{D(u)}$ for every $u \in \mathit{dom}(D)$, and
- for any $p \in \mathit{IP}$ and $\langle s, o \rangle \in \mathbf{I}_{\twoheadrightarrow}(p)$, $\langle \sigma(s), \sigma(o) \rangle \in \mathit{IEXT}(\sigma(p))$.

Satisfaction of all conditions up to and including 19 are established analogous to the ($\Leftarrow$) direction in the proof of Theorem 4. Notice that condition 20 is satisfied in $I$ by (v).

Consider a typed literal $t = (s, u') \in \mathcal{TL}$ and a datatype identifier $u \in \mathit{dom}(D)$. If $\mathit{IS}(u') = \mathit{IS}(u)$, then it must be the case that $D(u') = D(u)$, by construction of $I$. If $s \in L^{D(u')}$, then $(s, u')$ is a well-typed literal, and thus $\mathit{IL}((s, u')) = L2V^{D(u')} = L2V^{D(u)}$, by (iv). If $s \notin L^{D(u')}$, then $(s, u')$ is an ill-typed literal and $\mathit{IL}((s, u')) = (s, u') \notin \mathit{LV}$, because $\mathit{LV}$ does not contain ill-typed literals. Therefore, conditions 21 and 22 in Table 8 are satisfied.

Consequently, $I$ is an $x$-$D$-interpretation. We have that $I \models S$ and and thus $S$ is $x$-$D$-satisfiable.

The second part of the theorem follows from the observation that, for both directions, we have $I \not\models E$ iff $\mathbf{I} \not\models_{\mathsf{f}} \mathit{tr}(E)$. □

## References

Abiteboul, S., Hull, R., & Vianu, V. (1995). *Foundations of Databases.* Addison-Wesley.

Borgida, A. (1996). On the relative expressiveness of description logics and predicate logics. *Artificial Intelligence*, *82*(1–2), 353–367.

Brickley, D., & Guha, R. V. (2004). RDF vocabulary description language 1.0: RDF schema. Recommendation 10 February 2004, W3C.

Calvanese, D., Giacomo, G. D., Lembo, D., Lenzerini, M., & Rosati, R. (2007). Tractable reasoning and efficient query answering in description logics: the dl-lite family. *Journal of Automated Reasoning*, *39*, 385–429.

Dantsin, E., Eiter, T., Gottlob, G., & Voronkov, A. (2001). Complexity and expressive power of logic programming. *ACM Computing Surveys (CSUR)*, *33*(3), 374–425.

de Bruijn, J., Franconi, E., & Tessaris, S. (2005). Logical reconstruction of normative RDF. In *Proceedings of the Workshop OWL: Experiences and Directions (OWLED-2005).*

de Bruijn, J., & Heymans, S. (2007). Logical foundations of (e)RDF(S): Complexity and reasoning. In *Proceedings of the 6th International Semantic Web Conference (ISWC2007)*, pp. 86–99. Springer.

de Bruijn, J., & Heymans, S. (2008). On the relationship between description logic-based and f-logic-based ontologies. *Fundamenta Informaticae*, *82*(3), 213–236.

Fitting, M. (1996). *First Order Logic and Automated Theorem Proving (second edition).* Springer.

Gary, M. R., & Johnson, D. S. (1979). *Computers and Intractability – A Guide to the Theory of NP-Completeness.* W.H. Freeman and Company, New York, NY, USA.

Gutierrez, C., Hurtado, C., & Mendelzon, A. O. (2004). Foundations of semantic web databases. In *Proceedings of the 23rd ACM Symposium on Principles of Database Systems (PODS2004)*, pp. 95–106. ACM Press.

Gutierrez, C., Hurtado, C. A., Mendelzon, A. O., & Pérez, J. (2010). Foundations of semantic web databases. *Journal of Computer and System Sciences.* In Press.

Hayes, P. (2004). RDF semantics. Recommendation 10 February 2004, W3C.

Jones, N. D., & Laaser, W. T. (1974). Complete problems for deterministic polynomial time. In *Proceedings of the 6th Annual ACM Symposium on Theory of Computing (STOC1974)*, pp. 40–46, Seattle, Washington, USA. ACM Press.

Kifer, M., Lausen, G., & Wu, J. (1995). Logical foundations of object-oriented and frame-based languages. *Journal of the ACM*, *42*(4), 741–843.

Klyne, G., & Carroll, J. J. (2004). Resource description framework (RDF): Concepts and abstract syntax. Recommendation 10 February 2004, W3C.

Motik, B., Grau, B. C., Horrocks, I., Wu, Z., Fokoue, A., & Lutz, C. (2009a). OWL 2 web ontology language profiles. Recommendation 27 October 2009, W3C.

Motik, B., Patel-Schneider, P. F., & Parsia, B. (2009b). OWL 2 web ontology language structural specification and functional-style syntax. Recommendation 27 October 2009, W3C.

Muñoz, S., Pérez, J., & Gutierrez, C. (2009). Simple and efficient minimal RDFS. *Journal of Web Semantics*, *7*(3), 220–234.

Papadimitriou, C. H. (1994). *Computational Complexity*. Addison Wesley.

Patel-Schneider, P. F., Hayes, P., & Horrocks, I. (2004). OWL web ontology language semantics and abstract syntax. Recommendation 10 February 2004, W3C.

Peterson, D., Gao, S., Malhotra, A., Sperberg-McQueen, C. M., & Thompson, H. S. (2009). W3C XML schema definition language (XSD) 1.1 part 2: Datatypes. Working draft 3 December 2009, W3C.

RIF Working Group (2010a). RIF basic logic dialect. Recommendation 22 June 2010, W3C.

RIF Working Group (2010b). RIF RDF and OWL compatibility. Recommendation 22 June 2010, W3C.

ter Horst, H. J. (2005). Completeness, decidability and complexity of entailment for RDF schema and a semantic extension involving the OWL vocabulary. *Journal of Web Semantics*, *3*(2–3), 79–115.

Yang, G., Kifer, M., & Zhao, C. (2003). FLORA-2: A rule-based knowledge representation and inference infrastructure for the semantic web. In *Proceedings of the Second International Conference on Ontologies, Databases and Applications of Semantics (ODBASE2003)*. Springer.